\documentclass[10pt,conference]{IEEEtran}
\IEEEoverridecommandlockouts
\usepackage{cite}
\usepackage{amsmath,amssymb,amsfonts}
\usepackage{graphicx}
\usepackage{textcomp}
\usepackage{xcolor}
\usepackage{booktabs}
\usepackage{makecell}
\usepackage{multirow}
\usepackage[hyphens]{url}
\usepackage{hyperref}
\usepackage[T1,OT1]{fontenc}
\usepackage{listings}
\usepackage{upquote}
\usepackage{flushend}

\newcommand{\Description}[1]{}

\makeatletter
\newcommand{\compactsubsection}{%
  \@startsection{subsection}{2}{\z@}{1ex}%
  {0.7ex}{\normalfont\normalsize\itshape}}
\makeatother

\hypersetup{
  pdftitle={GSPIM: A Temporal-Locality-Aware 4D Gaussian Splatting Accelerator with Polymorphic PIM and Multi-Frame Rendering Dataflow},
  pdfauthor={Zirui Ma, Zhihua Fan, Haibin Wu, Dongrui Fan, Xiaochun Ye, Wenming Li},
  pdfsubject={Author manuscript},
  colorlinks=false,
  pdfborder={0 0 1},
  citebordercolor={0 0.55 0},
  linkbordercolor={0.8 0 0},
  urlbordercolor={0 0.55 0.75}
}

\title{GSPIM: A Temporal-Locality-Aware 4D Gaussian Splatting Accelerator with Polymorphic PIM and Multi-Frame Rendering Dataflow%
  \thanks{This work was supported by National Key R\&D Program of China (Grant No.2023YFB4503500), National Natural Science Foundation of China (Grant No.62502498, 62672477), Innovation Funding of \mbox{ICT, CAS (Grant No. E661120).}}}

\newcommand\gspimauthors{%
  Zirui Ma\textsuperscript{1,2}, Zhihua Fan\textsuperscript{1,2,*},
  Haibin Wu\textsuperscript{1,2}, Dongrui Fan\textsuperscript{3},
  Xiaochun Ye\textsuperscript{1,2}, and Wenming Li\textsuperscript{1,2,*}}
\newcommand\gspimaffiliation{%
  \textsuperscript{1}Institute of Computing Technology, Chinese Academy of Sciences, Beijing, China \\
  \textsuperscript{2}University of Chinese Academy of Sciences, Beijing, China \\
  \textsuperscript{3}Ricore IC Technologies Ltd., Beijing, China}
\newcommand\gspimemail{%
  \{mazirui24s, fanzhihua, wuhaibin, yexiaochun, liwenming\}@ict.ac.cn; fandr@smart-core.cn \\
  \textsuperscript{*}Corresponding authors: Zhihua Fan and Wenming Li.}

\author{%
  \IEEEauthorblockN{\gspimauthors{}}
  \IEEEauthorblockA{%
    \gspimaffiliation{} \\
    \gspimemail{}}
}

\begin{document}
\maketitle
\thispagestyle{plain}
\pagestyle{plain}

% Consistent separation between single-column top floats and body text.
\setlength{\textfloatsep}{12pt plus 1pt minus 2pt}

%%%%%%%%%%%%%%%%%%%%%%%%%%%%%%%%%%%%%%%%
%%%%%%%% -- PAPER CONTENT STARTS -- %%%%%%%%%

\begin{abstract}
4D Gaussian Splatting (4DGS) enables high-quality real-time novel-view synthesis for dynamic scenes, supporting free-viewpoint video, mobile XR, and autonomous systems at the edge.
Methods using explicit 4D distributions, keyframe interpolation, or anchor binding share two GPU data-movement bottlenecks.
Temporal filtering follows model record loading, wasting LPDDR5 bandwidth on inactive data.
Frame-isolated execution also overlooks highly overlapping active sets and depth relationships across adjacent frames, reloading shared attributes when working sets exceed on-chip cache capacity.
To address these bottlenecks, we present GSPIM, a temporal-locality-aware 4DGS accelerator for PIM-enabled edge devices.
Polymorphic PIM adapts a shared near-memory datapath to representation-specific field layouts and selection procedures.
It filters and compacts valid data into contiguous inputs before the GPU accesses complete attributes, reducing ineffective transfers.
Multi-Frame Rendering Dataflow reuses active attributes and stable depth relationships within a temporal window, reducing repeated work across frames.
Hybrid Dataflow Scheduling overlaps near-memory data preparation with GPU rendering to reduce waiting.
Evaluation combines a GPU--PIM model calibrated to Jetson AGX Orin 64 GB with a 1z DRAM physical implementation of the added circuits.
Across three 4DGS methods and six dynamic scenes, GSPIM achieves a 5.36$\times$ geometric-mean speedup and a 4.99$\times$ geometric-mean energy-efficiency improvement over a fully optimized frame-by-frame GPU baseline.
The maximum scene-level PSNR difference is 0.006 dB, and the added physical footprint is 18.7172 mm$^2$ across 32 LPDDR5 dies.
\end{abstract}

\section{Introduction}
\label{sec:introduction}

Novel-view synthesis for dynamic scenes~\cite{kplanes,hexplane,yang4dgs,4drotorgs,wu4dgs,stgfs} reconstructs time-varying 3D scenes from multi-view videos and generates high-quality images at arbitrary times and viewpoints, providing real-time visual content for free-viewpoint video~\cite{queen}, mobile XR~\cite{gazevis}, volumetric telepresence~\cite{realtime3dhr}, and autonomous systems~\cite{desiregs}. 3D Gaussian Splatting (3DGS)~\cite{3dgs,scaffoldgs} represents a static scene with an explicit set of Gaussians and achieves high-quality real-time rendering through projection, sorting, and forward compositing. 4D Gaussian Splatting (4DGS)~\cite{yang4dgs,4dgs1k,ex4dgs,anchored4dgs} further describes temporal scene variation, allowing the same dynamic representation to generate the corresponding 3D Gaussians at consecutive timestamps. Explicit 4D distributions jointly model spacetime, keyframe interpolation generates continuous motion from sparse samples, and anchors share temporal parameters to reduce redundancy. They determine activity from per-record temporal fields before loading geometry and appearance for per-frame rendering. Edge GPUs provide a programmable platform, but limited LPDDR5 bandwidth and on-chip cache capacity make large-scale Gaussian data movement a primary constraint on real-time inference.

Prior work reduces the software overhead of 4DGS through representation compression, primitive pruning, and CUDA kernel optimization~\cite{yang4dgs,mega,4dgs1k}, while specialized accelerators build generation and rendering datapaths around specific temporal representations~\cite{3dgaucim,thudgs}. Existing specialized accelerators exploit spatial or inter-frame correlation but do not organize shared source attributes into multi-frame batches. Our analysis of frame-by-frame GPU execution reveals two data-access problems. First, temporal filtering occurs after data access, leaving 73.3\% to 86.8\% of LPDDR5 transfers unused by subsequent computation. Second, adjacent frames have highly overlapping active sets and depth relationships, yet frame-isolated execution repeatedly loads attributes shared across frames. These problems arise from late temporal decisions and a frame-by-frame GPU dataflow that does not organize shared attributes for window-level reuse.

\begin{figure*}[!t]
  \centering
  \includegraphics[width=\textwidth]{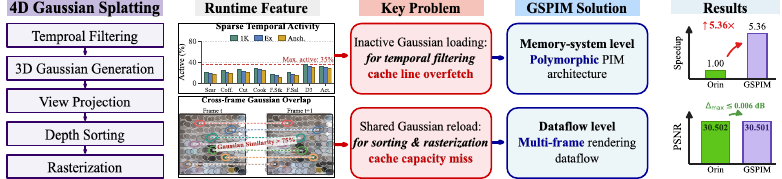}
  \caption{4DGS pipeline, memory-access bottlenecks, and GSPIM overview.}
  \label{fig:intro}
  \Description{An overview connecting the 4DGS pipeline and its temporal characteristics to the near-memory and multi-frame solutions in GSPIM.}
\end{figure*}

To address these data-movement bottlenecks, we present GSPIM, an edge 4DGS accelerator that exploits temporal locality to improve throughput (Fig.~\ref{fig:intro}).
Polymorphic PIM (PPIM) maps local decisions for different temporal representations to a programmable bank-side datapath.
Its selection results drive bank- and die-level compaction, removing inactive attributes and packing valid data contiguously before GPU access to reduce ineffective LPDDR5 transfers.
Multi-Frame Rendering Dataflow groups shared attributes into batches that fit in L2.
It reuses attributes and sorting results for stable primitives across frames in each window, reducing repeated loads and per-frame sorting.
Hybrid Dataflow Scheduling overlaps near-memory data preparation with GPU computation across windows and batches, reducing resource idle time from serial stage execution.
This work makes the following contributions.

\textbf{(i)} We propose Polymorphic PIM. PPIM maps local decisions from different 4DGS representations and dataflow stages onto a shared fixed-point datapath. It produces a unified selection result that drives bank- and die-level data compaction, excluding ineffective attributes before they reach the GPU.

\textbf{(ii)} We propose a Multi-Frame Rendering Dataflow and its Hybrid Dataflow Scheduling. The dataflow reuses active attributes and stable depth relationships within temporal windows and sorts only primitives with substantial depth variation per frame. Double buffering sustains concurrent near-memory batch preparation and GPU multi-frame rendering across consecutive windows and batches.

\textbf{(iii)} We implement the near-memory hardware of GSPIM in Chisel RTL and evaluate it on a GPU--PIM simulation platform calibrated to the Jetson AGX Orin 64 GB.\footnote{\scriptsize Our open-source simulator: \url{https://github.com/MAdrid1011/GSPIM}.} Across three 4DGS methods and six scenes, GSPIM achieves a 5.36$\times$ geometric-mean speedup and a 4.99$\times$ geometric-mean improvement in energy efficiency over a fully optimized frame-by-frame GPU path. The maximum scene-level PSNR difference is 0.006 dB, while SSIM differs from the baseline by at most 0.0001. In 1z DRAM, the added circuits have a physical footprint of 18.7172 mm$^2$ and active power of 3.8363 W across 32 dies.

\section{Background and Motivation}

\subsection{4D Gaussian Splatting}\label{sec:4dgs-pipeline}

3D Gaussian Splatting (3DGS)~\cite{3dgs} represents a static scene as Gaussian primitives with a 3D center, covariance, opacity, and appearance. Rendering applies projection, depth sorting, and forward compositing. 4D Gaussian Splatting (4DGS) adds temporal variation to generate 3D Gaussians across timestamps. Efficient explicit representations use 4D distributions, keyframe interpolation, or anchor binding. Explicit 4D Gaussians~\cite{yang4dgs,mega,4dgs1k} jointly encode spacetime in 4D means and covariances, supporting temporal slicing. Keyframe interpolation~\cite{ex4dgs} generates continuous motion from sparse attribute samples. Anchors~\cite{anchored4dgs} share temporal parameters within valid intervals to generate or select bound Gaussians.

Near-bank temporal filtering targets representations that explicitly store local temporal fields in Gaussian or anchor records and uses these fields to determine window activity. As shown on the left of Figure~\ref{fig:intro}, temporal filtering selects active records, and 3D Gaussian generation loads their geometric and appearance attributes to construct standard 3D Gaussians at each timestamp. View projection computes screen positions and 2D footprints, depth sorting determines compositing order, and rasterization accumulates pixel color and transmittance in that order to produce the output image. Field layouts, decision procedures, and Gaussian counts per record differ, but activity determines attribute access. GPU-side temporal filtering incurs ineffective transfers before selection, while frame-by-frame sorting and rasterization reload shared attributes. Highly overlapping active sets and depth relationships across frames offer attribute and sorting reuse opportunities.

\subsection{Processing in Memory}\label{sec:pim-background}

Edge GPUs commonly share LPDDR memory with CPUs and media engines, making data movement costly under limited bandwidth and cache capacity. Processing in Memory (PIM)~\cite{pimsys,imprint,pimscalability,centralizedpim,aquabolt-xl,paise} adds lightweight computation to bank peripherals or intra-die datapaths, allowing filtering, reduction, and data reorganization before the GPU accesses complete data. Prior studies~\cite{pimdl,neupims,specpim,pimsurvey} show that memory-intensive workloads benefit from reduced data movement and energy use.

\begin{figure}[!t]
  \centering
  \includegraphics[width=\columnwidth]{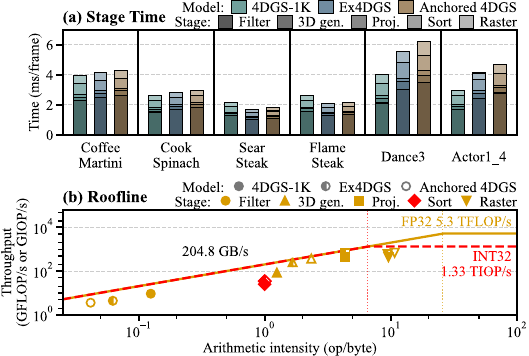}
  \caption{Baseline stage time and roofline on AGX Orin.}
  \label{fig:baseline-pipeline}
  \Description{Baseline execution time broken down into five stages for three 4DGS methods and six scenes, followed by fifteen stage roofline points compared with the corresponding FP32 or INT32 ceiling.}
\end{figure}

The temporal processing of 4DGS matches these characteristics. Each of millions of model records needs only a few local temporal fields to produce a compact activity result controlling access to much larger geometric and appearance attributes. Despite differing field layouts and temporal computations, activity decisions use bounded arithmetic or interval tests over fixed ranges of local fields. Figure~\ref{fig:baseline-pipeline}(a) shows that temporal filtering accounts for 52.4\% to 63.3\% of baseline execution time on Jetson AGX Orin, while sorting and rasterization together account for 26.5\% to 34.8\%. Their arithmetic intensities fall below the corresponding roofline ridge points in Figure~\ref{fig:baseline-pipeline}(b), with temporal filtering lowest. Near-memory compaction and reorganization use these decisions to exclude inactive records before attribute transfer and provide contiguous inputs for sorting and rendering, reducing wasted transfers and scattered accesses during both initial filtering and subsequent data preparation. Configurable field access and decisions support different 4DGS representations on a shared PIM datapath, while the GPU generates their standard 3D Gaussians and performs projection, sorting, and rasterization.

\subsection{Challenges in Accelerating Edge 4DGS}

We analyze 4DGS-1K~\cite{4dgs1k}, Ex4DGS~\cite{ex4dgs}, and Anchored 4DGS~\cite{anchored4dgs} on a Jetson AGX Orin 64 GB. These methods cover explicit 4D Gaussians, keyframe interpolation, and anchor binding, respectively. We use N3V~\cite{n3v} and EasyVolcap~\cite{easyvolcap}, which include local object motion, large nonrigid motion, and varying degrees of occlusion change. All three methods use fully optimized CUDA paths under the same LPDDR5 and 4 MB L2 configuration, exposing two common limits to efficient edge 4DGS inference.

\paragraph{\textbf{Challenge 1: Ineffective loads caused by GPU-side temporal filtering}}

\begin{figure}[!t]
  \centering
  \includegraphics[width=\columnwidth]{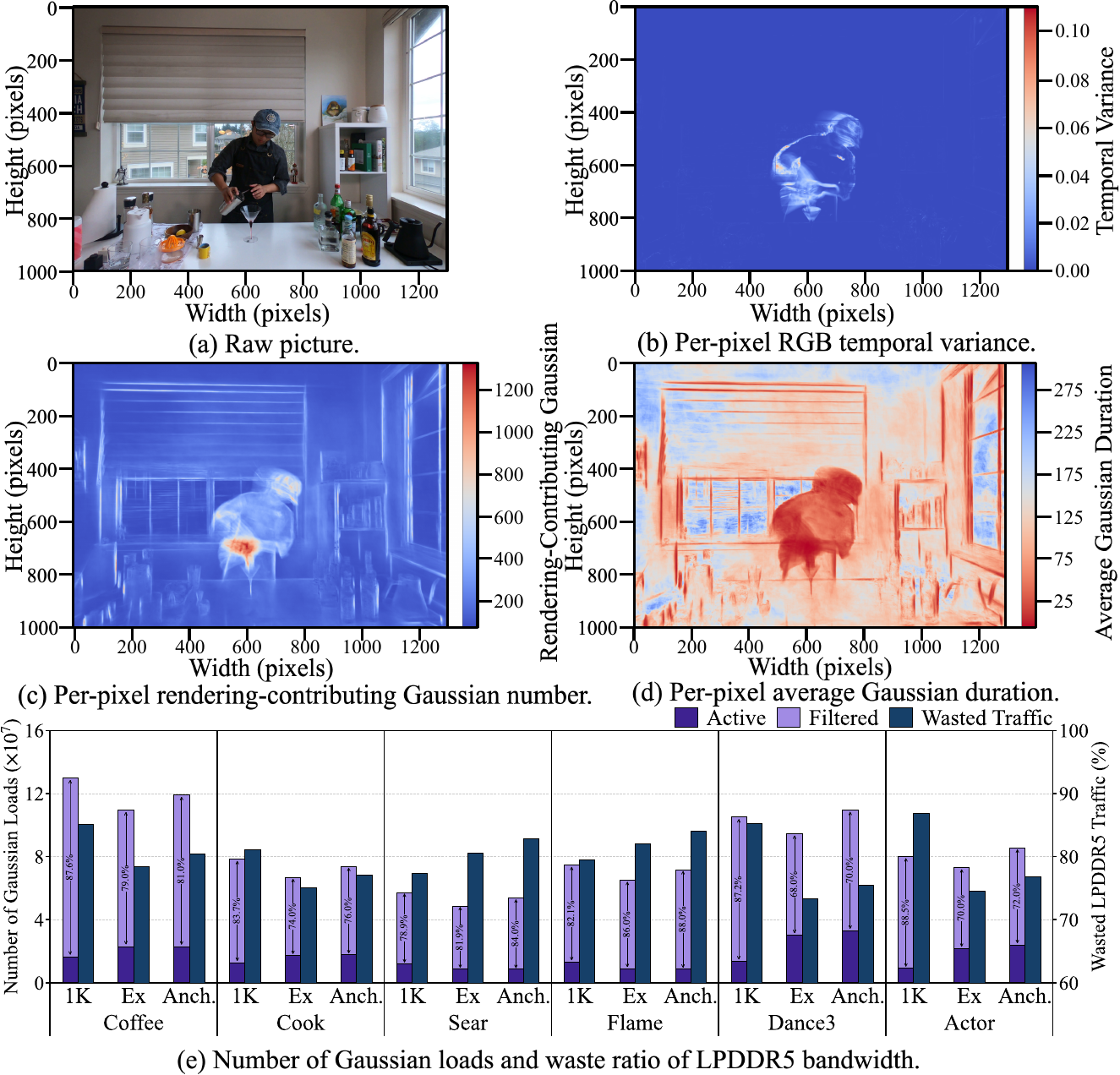}
  \caption{Temporal sparsity and ineffective Gaussian loading.}
  \label{fig:challenge1}
  \Description{Measurements showing temporal sparsity and ineffective Gaussian loads across three 4DGS workloads and multiple dynamic scenes.}
\end{figure}

Dynamic scenes exhibit inherent temporal sparsity. Foreground objects usually appear over a limited interval of the complete sequence, while static backgrounds remain active for long periods. Consequently, only a subset of model records produces valid Gaussians in any temporal window. Figure~\ref{fig:challenge1} shows that more than 70\% of the primitives in 4DGS-1K are discarded after temporal filtering, resulting in 77.4\% to 86.8\% ineffective memory transfers. In Ex4DGS, keyframe intervals outside the current window cause 73.3\% to 82.0\% ineffective transfers. In Anchored 4DGS, each inactive anchor controls multiple Gaussians, yet its activity is known only after fetching anchor parameters and binding relations, causing 75.5\% to 84.0\% ineffective transfers.

Existing GPU dataflows first read temporal fields from LPDDR5 and then determine activity. Active and inactive records are interleaved by global ID across attribute arrays, so cache-line accesses also fetch adjacent inactive fields and consume limited L2 capacity. This process wastes both the temporal fields discarded after the decision and the geometric and appearance attributes unnecessarily prefetched for the current window. Temporal fields occupy only 8 B to 52 B per record across the three representations, but determine whether as many as 584 B of subsequent attributes must be accessed. Across all three representations, GPU-side activity decisions make temporal filtering a common LPDDR5 bottleneck because ineffective accesses have already occurred.

\paragraph{\textbf{Challenge 2: Frame-isolated rendering causes repeated loading of reusable Gaussians}}

\begin{figure}[!t]
  \centering
  \includegraphics[width=\columnwidth]{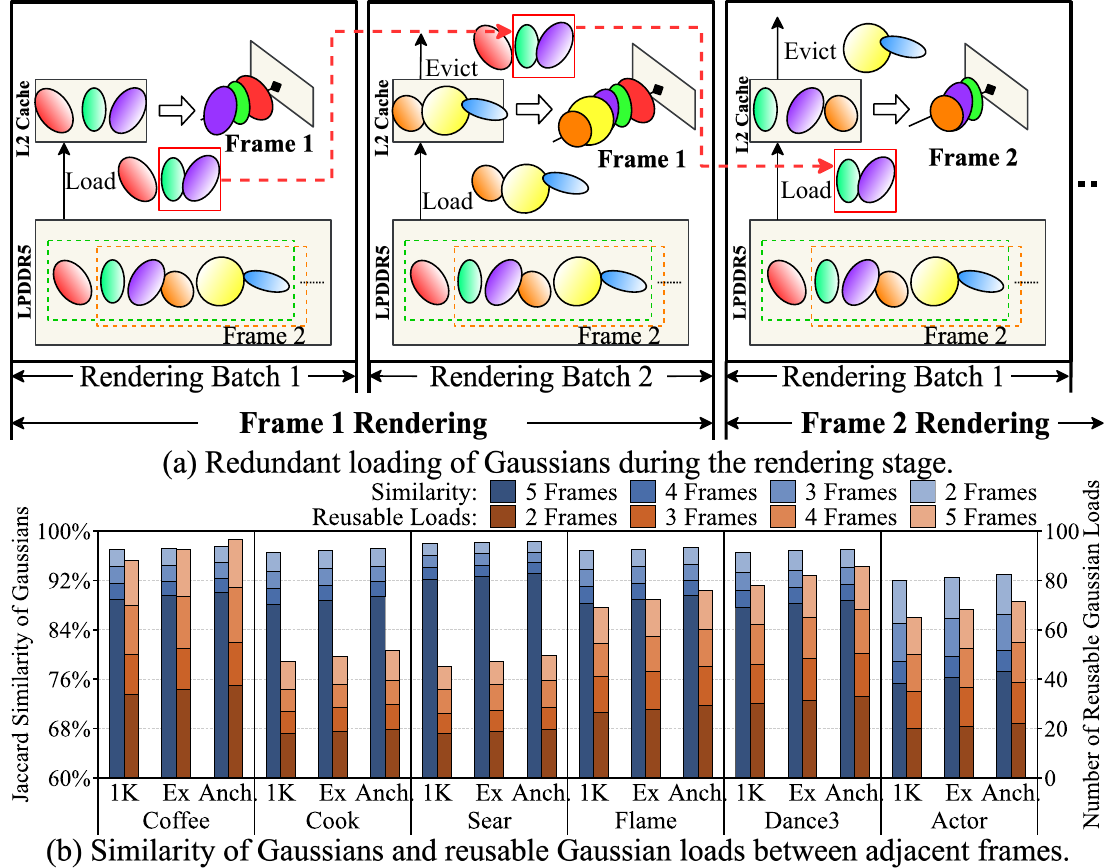}
  \caption{Cross-frame reuse under frame-isolated rendering.}
  \label{fig:challenge2}
  \Description{An illustration and measurements of repeated Gaussian loading under frame-isolated execution.}
\end{figure}

The temporal continuity of dynamic scenes gives adjacent frames highly overlapping active sets and depth relationships~\cite{temporalcoherence}. Figure~\ref{fig:challenge2} shows that active sets retain high Jaccard similarity across all scenes within temporal windows of two to five frames. Across all evaluated methods and scenes, five-frame active-set similarity remains at least 75.3\%. Actor1\_4 forms the lowest-overlap case because its large nonrigid motion and occlusion changes replace more Gaussians within a window. These Gaussians share model attributes, while their per-frame geometry and depth relations vary smoothly enough to remain reusable across consecutive frames in the same temporal window.

Frame-isolated 4DGS dataflows reaccess active primitives, generate projections, and construct a depth order per frame. The per-frame active working set commonly exceeds the 4 MB AGX Orin L2 cache, while projection, sorting, and rasterization intermediates consume additional capacity. Shared attributes are thus evicted before the next frame uses them. As the window grows, unique Gaussians increase slowly, but cumulative multi-frame accesses grow approximately with frame count. Limited edge cache capacity prevents temporal overlap from yielding reuse, leaving frame-isolated execution to repeat LPDDR5 accesses and attribute loads.

\begin{figure*}[!t]
  \centering
  \includegraphics[width=\textwidth]{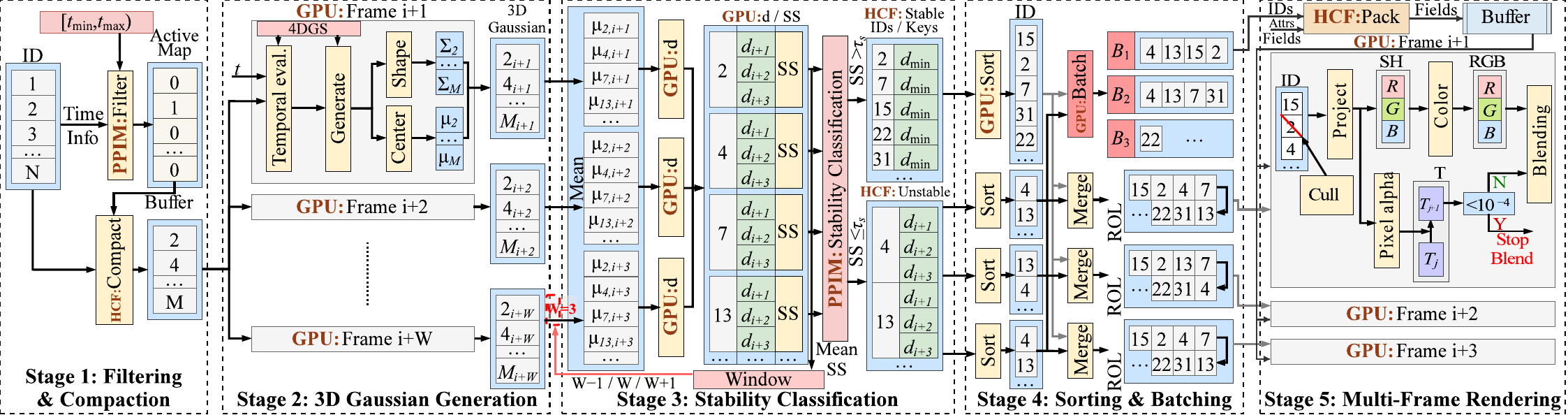}
  \caption{GSPIM multi-frame rendering dataflow.}
  \label{fig:dataflow}
  \Description{The five-stage GSPIM dataflow with GPU, PPIM, and HCF execution assignments and the transferred Active Map, Active Buffer, depth fields, classification metadata, batch indices, and packed attributes.}
\end{figure*}

\subsection{Our Approach}

GSPIM exploits temporal locality in edge 4DGS, sharing window execution across GPU and PIM through Polymorphic PIM and Multi-Frame Rendering Dataflow.

\paragraph{\textbf{Polymorphic PIM}}

Polymorphic PIM maps the field layouts and decision procedures of different 4DGS representations onto a shared near-memory datapath and produces a uniform selection result before geometric and appearance attributes are accessed. The result drives a separate two-level path that compacts sparse valid data into contiguous stage inputs, reducing scattered access and cache-line overfetch.

\paragraph{\textbf{Multi-Frame Rendering Dataflow}}

Multi-Frame Rendering Dataflow groups consecutive frames into a temporal window and shared attributes into batches, reducing repeated loads. The GPU reuses each batch across frames while preserving per-frame depth order, as near-memory execution prepares the next batch without increasing GPU cache pressure.

\section{GSPIM Dataflow}\label{sec:gspim-dataflow}

GSPIM adopts a Multi-Frame Rendering Dataflow that processes consecutive frames within a temporal window in batch-major order to reuse active Gaussian primitives (Fig.~\ref{fig:dataflow}).
PPIM makes activity and stability decisions, the Hierarchical Compaction Fabric (HCF) compacts, reorganizes, and packs data, and the GPU generates geometry, computes depths, sorts, merges, batches, and renders.
For temporally sparse 4DGS representations with different temporal parameterizations, Stage 1 configures activity computations from local temporal fields, while later stages process standard 3D Gaussian parameters.

\subsection{Stage 1: Temporal Filtering and Compaction}

Stage 1 performs temporal filtering and compaction over the current temporal window $[t_{\min},t_{\max})$, where $t_{\min}$ and $t_{\max}$ denote the start and end of the window. To address the sparse, interleaved activity in Challenge 1, PPIM reads local temporal information to determine window activity and produces a one-bit result per primitive as the Active Map. HCF compacts marked primitives into a contiguous Active Buffer. Stage 2 generates all frame-specific 3D Gaussians from this buffer.

\subsection{Stage 2: Multi-Frame 3D Gaussian Generation}

Stage 2 uses the current temporal window as a GPU batch and generates the corresponding 3D Gaussians for all frames in parallel. The GPU configures the computation for each 4DGS representation. Frame channels reuse the active data in the Active Buffer and use their corresponding frame timestamps to compute the 3D mean $\mu^{(3)}$ and covariance $\Sigma^{(3)}$ of every active primitive. This stage converts representation-specific temporal variation into a uniform sequence of multi-frame 3D Gaussians, allowing subsequent stages to process a common data format and providing Stage 3 with the 3D centers of all frames in the window.

\subsection{Stage 3: Depth Stability Score Classification}

Stage 3 first transforms the 3D centers in the window into camera coordinates on the GPU. For camera-space depth $z_i$ and rendering bounds $z_{\mathrm{near}}$ and $z_{\mathrm{far}}$, GSPIM forms the dimensionless sequence $\mathcal{D}=\{d_1,\ldots,d_W\}$, where $d_i=(z_i-z_{\mathrm{near}})/(z_{\mathrm{far}}-z_{\mathrm{near}})$. To preserve visibility, conventional 4DGS methods commonly perform global depth sorting over all primitives in every frame, incurring substantial repeated sorting overhead. Depth changes across adjacent frames are usually smooth in natural scenes, while only a small subset of primitives experiences substantial changes because of fast motion or sudden occlusion. Based on this observation, GSPIM uses a depth stability score $SS$ to describe the depth stability of each primitive within a window and maps its average depth variation into a normalized range through exponential decay:
\begin{equation}
SS=\exp\left\{-\frac{1}{W-1}\sum_{i=1}^{W-1}\left|d_{i+1}-d_i\right|\right\}.
\label{eq:stability-score}
\end{equation}

A high $SS$ indicates stable cross-frame depth, while a low value indicates substantial variation. The GPU reduces the average depth change while generating the depth sequence, and PPIM produces a classification bit through a threshold comparison equivalent to $SS>\tau_s$, where $\tau_s$ is the stability threshold. According to the two values of this bit, HCF organizes the indices and sorting keys of the two primitive classes into separate contiguous sequences, allowing the Stage 4 GPU to read both sorting inputs contiguously. Source attributes and per-frame geometry remain in their original regions. The mean $SS$ over all window primitives reflects overall dynamics. A mean below the window-adjustment threshold $\tau_w$ indicates greater depth variation and a larger unstable proportion, so the system shortens the next window by one frame to limit reuse under fast changes. Otherwise, it adds one frame within the two-to-five-frame range, as most primitives retain stable cross-frame depth relationships and share classification and sorting across more frames.

\subsection{Stage 4: Sorting and Batch Scheduling}

\paragraph{\textbf{Stability-aware differentiated sorting}}

Prior work~\cite{flashgs} shows that depth sorting~\cite{hybridradixsort} is a major cost in the Gaussian Splatting pipeline. Our measurements show that $SS$ follows the same trend as cross-frame position retention, with high-$SS$ primitives preserving their depth ranks more consistently than low-$SS$ primitives. The stable set $\mathcal{P}_s$ and unstable set $\mathcal{P}_u$ produced by threshold classification therefore have different sorting requirements. GSPIM exploits this distinction through stability-aware differentiated sorting, assigning each set a separate key, frequency, and algorithm.

Let the stable and unstable sets contain $N_s$ and $N_u$ primitives, respectively. For $\mathcal{P}_s$, the relative occlusion order remains nearly unchanged within a window, so one radix sort using the per-window minimum depth $d_{\min}$ serves all $W$ frames and amortizes the $O(N_s)$ fixed-width sorting cost over the complete window. To preserve visibility for $\mathcal{P}_u$, GSPIM still performs per-frame local radix sorting. Most primitives remain stable within a window in our experiments, so per-frame sorting applies only to the smaller unstable set and costs $O(N_u)$ per frame. The GPU uses parallel merge-path~\cite{mergepath} to merge the shared stable sequence with per-frame unstable sequences into render order lists (ROLs) for all $W$ frames.

\paragraph{\textbf{Batching}}

To preserve both cross-frame attribute reuse and per-frame depth order under limited L2 capacity, GSPIM uses the window-level order of stable primitives as a shared batch boundary across frames. The stable portions of the per-frame ROLs preserve a common relative order, and their differences primarily arise from where unstable primitives are inserted. A shared boundary can therefore align the same stable primitives within one batch. The GPU extends the current batch along the stable sequence and gathers primitives between the same boundaries in every frame ROL. Shared attributes are stored once, while local orders retain per-frame differences. Based on the actual space occupied by attributes, local indices, and rasterization intermediates, the system selects the largest stable interval that fits within the effective L2 capacity. One load of the same attribute batch consequently serves multiple frames in the window, while every frame still performs depth-ordered compositing according to its original ROL.

\subsection{Stage 5: Batch Multi-Frame Rendering}

Stage 5 renders Stage 4 batches through the standard 3DGS rasterization pipeline~\cite{3dgs}. The GPU loads shared attributes of the current batch into L2 for reuse across window frames, combining them with per-frame dynamic fields for 2D projection, tile assignment, and forward compositing. The GPU follows per-frame local batch orders, retaining color and transmittance across batches to preserve the original ROL in the concatenated sequence. While the GPU reuses these attributes to render the current batch, HCF packs the shared attributes and per-frame fields for the next group of indices into another Batch Buffer. The next batch stays in main memory until the switch, outside L2. After batch completion, the Batch Buffers swap roles, pipelining multi-frame rendering with preparation of the next attribute batch.

\section{GSPIM Architecture}

\begin{figure}[!t]
  \centering
  \includegraphics[width=\columnwidth]{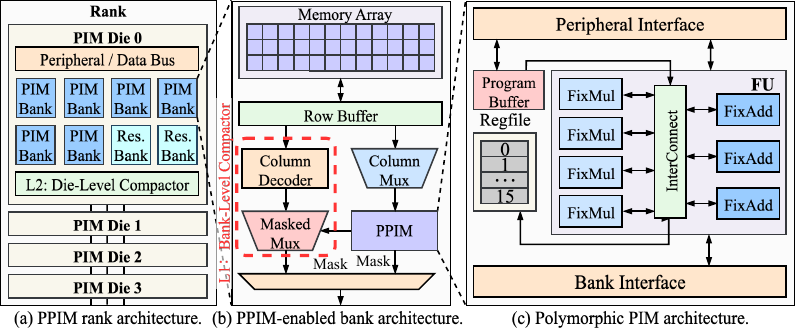}
  \caption{GSPIM architecture.}
  \label{fig:arch}
  \Description{The GSPIM rank, a PPIM-enabled bank, and the programmable fixed-point PPIM datapath.}
\end{figure}

As shown in Fig.~\ref{fig:arch}, GSPIM integrates Polymorphic PIM (PPIM) and a Hierarchical Compaction Fabric (HCF) within a rank. PPIM resides in PIM banks and performs representation-specific local decisions. HCF filters data within PIM banks according to the masks produced by PPIM and gathers results from multiple banks into reserved banks (Res. Banks).

\subsection{Polymorphic PIM}\label{sec:polymorphic-pim}

Different temporal parameters and associations encode 4DGS scene variation, but a few local fields determine window activity and cross-frame depth stability. GSPIM places PPIM beside each PIM-bank row buffer to support these decisions through the shared near-memory computation path in Fig.~\ref{fig:arch}(c). The column multiplexer reads the required local fields from the row buffer of its bank. Floating-point decision fields pass through a bank-side FP32 Converter before entering PPIM registers. It aligns significands by exponent and rounds $x$ to $q=\operatorname{round}(x2^{20})$, representing $q2^{-20}$. The bank program buffer instructs shared fixed-point units through the interconnect to run short programs and generate block-level masks. Decision-field reads, conversion, computation, and mask generation remain bank-local before HCF filters and gathers selected data. Programs derived from static representation layouts are written into every PIM-bank program buffer at model loading. PPIM combines 20 fractional bits, giving a quantization step below $10^{-6}$, with six integer bits including the sign, yielding signed 26-bit operands over $[-32,32)$. Conversion error is at most $2^{-21}$ without overflow. Program constants share this scale. Multiplication retains full 52-bit products until register writeback restores 20 fractional bits.

To translate PPIM decisions directly into fewer data reads, GSPIM modifies the original bank access path into a mask-controlled column-selection path. The column decoder selects fields from the row buffer according to the access address, and the masked multiplexer retains fields selected by the PPIM mask before forwarding them to the bank interface. Regular GPU accesses use a fully valid mask, while data compaction uses the S1 window-activity mask or the S3 depth-stability mask to filter target fields before they leave the bank. For results within one output least-significant bit of a threshold, S1 retains the Gaussian and S3 assigns it to the unstable set for per-frame sorting, limiting quantization effects on filtering and classification. The computation path adapts to different decisions through programs, while the access path controls data output through a uniform mask.

\begin{figure}[!t]
  \centering
  \includegraphics[width=\columnwidth]{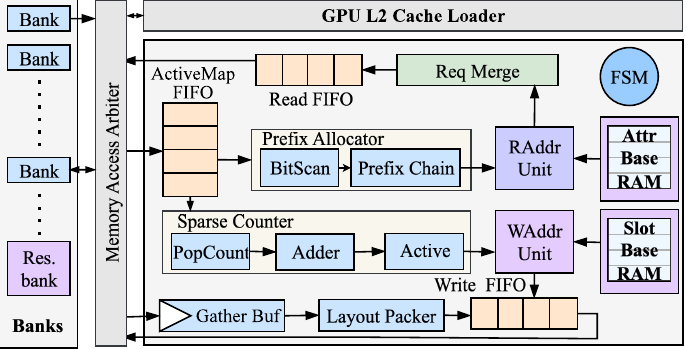}
  \caption{L2 die-level compactor.}
  \label{fig:hcf}
  \Description{The die-level HCF path for scanning selection maps, allocating destinations, gathering bank data, and writing compacted outputs.}
\end{figure}

\subsection{Hierarchical Compaction Fabric}\label{sec:hierarchical-compaction}

To convert distributed selection results into contiguous data, HCF uses an L1 Bank-Level Compactor in each PIM bank and an L2 Die-Level Compactor in each die. L1 filters row-buffer fields by mask. Figure~\ref{fig:hcf} shows L2 scanning the ActiveMap FIFO, allocating contiguous destinations, and gathering selected data across banks. The four Res. Banks share the Active Buffer, classification sequences, and ping-pong Batch Buffers according to their live ranges, with overflow redirected to GPU reorganization. The Memory Access Arbiter prioritizes refresh, then GPU over HCF for same-bank conflicts. Before all-bank refresh, the target rank stops new accesses, drains in-flight transfers, and pauses PPIM and HCF. PPIM retains its instruction position and registers, while HCF retains FIFO contents and transfer progress. After refresh, execution resumes once the required rows are reactivated under normal access timing.

\begin{table}[t]
  \centering
  \caption{List of GSPIM commands.}
  \label{tab:isa}
  \small
  \begin{tabular}{@{}c|c@{}}
    \hline
    \textbf{Command} & \textbf{Description} \\
    \hline
    \texttt{PIM\_SELECT} & Generate block-level masks with a PPIM program \\
    \texttt{PIM\_REORG} & Reorganize selected data into contiguous ranges \\
    \hline
  \end{tabular}
\end{table}

HCF reuses one path in S1, S3, and batch packing. S1 compacts window-active data into a contiguous Active Buffer so S2 skips inactive attributes. In S3, L2 allocates separate stable and unstable destinations, applying the classification mask and its inverse to form contiguous sorting metadata and narrow the S4 per-frame sorting access range. For each S4 batch, HCF gathers shared attributes and per-frame fields into a Batch Buffer for contiguous S5 GPU loads and within-window attribute reuse. These uses share selection scanning, address generation, gathering, and contiguous writeback, differing only in selection inputs and destination layouts.

\subsection{Data Mapping}\label{sec:data-mapping}

To allow a PPIM mask to directly locate fields accessed later, GSPIM organizes model records into fixed-size data blocks. At model loading, the runtime distributes blocks across PIM banks and allocates row regions for decision fields and their associated attributes or sorting metadata from a driver-managed physical memory pool, selecting a channel, die, and bank for each block. Field writes preserve original values, precision, Gaussian associations, and common record order. Field row bases and record strides determine the row and column positions of each record. The runtime constructs each DRAM burst address from these positions. With XOR interleaving, the physical bank-selection bits are the XOR of the target bank identifier and the corresponding interleaving term, preserving the selected bank as row and column positions change. Allocated physical pages remain fixed throughout block use, and layouts retain field row bases and record strides for reuse across windows.

S1 temporal fields are written in this layout at model loading, while the GPU writes S3 stability-decision fields and associated sorting metadata directly to preallocated positions. This shared record order allows the masked multiplexer to filter corresponding records using the mask generated from decision fields read by the column multiplexer. When one decision record corresponds to one Gaussian, its mask bit directly identifies the aligned attributes of that Gaussian. In representations where one temporal record is associated with multiple Gaussians, the model stores the start position and count of that Gaussian range within the same die, and L2 reads attributes in the corresponding range for each selected record. After S1 completes mapping and compaction, it generates an Active Index for every retained Gaussian. The multi-frame geometry from S2, classification metadata from S3, and batch selections from S4 retain this index, allowing later reorganization to locate fields regardless of how the source representation stores its records.

\section{GSPIM Runtime}

% The GSPIM runtime integrates memory-side PPIM and HCF with the CUDA rendering pipeline and uses a task interface to express decisions, data reorganization, and their dependencies on GPU computation. The runtime generates task descriptors, submits PIM commands, propagates completion events, and organizes window-level execution and batch-level data preparation. This section describes the task-offloading interface and Hybrid Dataflow Scheduling.

\subsection{Task Offloading Interface}\label{sec:task-offloading}

GSPIM separates GPU and PPIM compilation. S2--S5 use the CUDA toolchain, whereas representation-specific decisions in S1 and S3 compile into PPIM microcode for field access, fixed-point arithmetic, comparison, and mask generation. Each model package contains the activity program, temporal-field layout, attribute mapping, and shared stability program. The runtime loads this metadata through standard LPDDR5 \texttt{WRITE} commands and submits two die-level tasks through reserved control addresses to the on-die command frontend. \texttt{PIM\_SELECT} executes a PPIM program and records block masks. \texttt{PIM\_REORG} uses HCF to count selected data, allocate contiguous ranges, and perform S1 compaction, S3 classification reorganization, or batch packing. Per-die in-order submission allows reorganization to consume the preceding selection result, while anchor-based paths allocate output space from binding lengths. Standard \texttt{READ} commands retrieve task status maintained by the completion unit. Task submission and completion queries reuse the existing host LPDDR5 controller and PHY. Table~\ref{tab:isa} summarizes both tasks.

\begin{figure}[!t]
  \centering
  \includegraphics[width=\columnwidth]{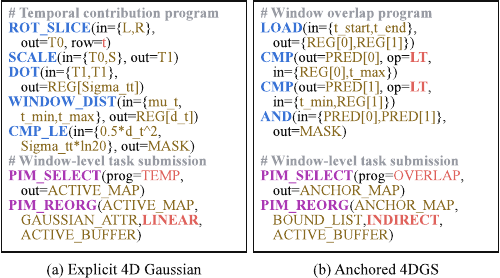}
  \caption{PPIM programming examples.}
  \label{fig:program}
  \Description{Two compact examples showing representation-specific PPIM programs and their common task-level submission interface.}
\end{figure}

For S1 and S3, the runtime submits \texttt{PIM\_SELECT} followed by \texttt{PIM\_REORG} for each die. Batch preparation directly submits \texttt{PIM\_REORG} using GPU selections. HCF exposes completed-block output ranges so the memory controller can release dependent GPU tasks, reporting the final output count when the task completes. The runtime writes source cache lines back before PIM execution and invalidates destination lines before GPU consumption. Figure~\ref{fig:program} shows two S1 programs, whose uppercase labels denote PPIM micro-operations rather than host commands. Explicit 4D Gaussian selection sums squared, per-dimension-scaled temporal rotation-matrix row entries to obtain temporal variance. Anchor-based selection tests temporal overlap and expands selected anchors via a binding table. Both use \texttt{PIM\_SELECT} to form masks and \texttt{PIM\_REORG} to compact Active Buffers.

\subsection{Hybrid Dataflow Scheduling}

To remove resource idleness caused by window-level stage serialization, GSPIM uses Hybrid Dataflow Scheduling according to cross-stage dependencies. For S1--S3, whose dependencies are confined to the same data block, a fine-grained block pipeline starts S2 for the first Active Buffer block as soon as HCF produces it, and uses available depth results to trigger S3 classification and reorganization. This policy reduces interstage synchronization to one data block. S4--S5 require a complete window to establish global depth order, so a coarse-grained batch pipeline preserves sorting boundaries while overlapping data preparation with rendering. After S3 determines the width of the next window, PPIM begins S1 selection for that window concurrently with S4 and S5 of the current window. The GPU, PPIM, and HCF thus run continuously at each stage's required granularity, reducing intra-window and cross-window resource waits.

\begin{figure}[!t]
  \centering
  \includegraphics[width=\columnwidth]{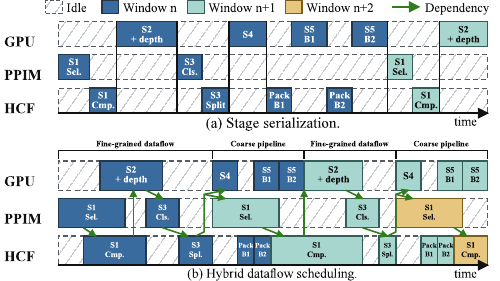}
  \caption{GSPIM hybrid dataflow scheduling.}
  \label{fig:pipeline}
  \Description{A timing diagram comparing serialized execution with hybrid block- and batch-level GPU--PIM pipelining.}
\end{figure}

Figure~\ref{fig:pipeline}(b) illustrates the steady execution rhythm formed by the hybrid dataflow. After the first window starts, each Active Buffer block triggers its corresponding S2 and S3 tasks, overlapping GPU consumption with subsequent selection and compaction. Once S3 adjusts the next window width to $W-1$, $W$, or $W+1$, GPU sorting and rendering for the current window primarily access Res. Banks, while S1 for the next window reads temporal fields from PIM banks. These accesses mainly target different banks, reducing same-bank contention. HCF processes batch packing and S1 compaction in dependency order and gathers selected data into Res. Banks. Its ActiveMap FIFO retains selection results produced early by PPIM and continues forming the Active Buffer when the reorganization path becomes available. Adjacent windows repeat this rhythm, keeping selection, data reorganization, and GPU computation continuously overlapped.

\section{Methodology}

\subsection{Workloads}

We evaluate three representative 4DGS workloads: 4DGS-1K (1K)~\cite{4dgs1k}, Ex4DGS (Ex)~\cite{ex4dgs}, and Anchored 4DGS (Anch.)~\cite{anchored4dgs}. They cover activity-mask filtering, keyframe interpolation, and anchor binding. N3V~\cite{n3v} and EasyVolcap~\cite{easyvolcap} span indoor and outdoor scenes with local, nonrigid, and occlusion changes. For a fully optimized GPU comparison, we connect representation-specific temporal computation and common Gaussian operators to a unified CUDA backend. CUB primitives implement filtering, prefix sums, and segmented radix sorting, and gsplat~\cite{gsplat} provides projection and rasterization. We tune thread configurations and batch sizes for each method on AGX Orin and verify frame-by-frame outputs against the original implementations. Sensitivity curves use representative smooth and nonrigid scenes, with selected thresholds verified across all workloads and datasets.

\begin{table}[t]
  \centering
  \caption{Hardware configuration of GSPIM.}
  \label{tab:platform}
  \resizebox{\columnwidth}{!}{%
  \renewcommand{\arraystretch}{1.12}
  \begin{tabular}{|l|c|l|c|}
    \hline
    \multicolumn{2}{|c|}{\textbf{GPU Configuration (AGX Orin)}} &
    \multicolumn{2}{c|}{\textbf{PIM-LPDDR5 Configuration}} \\ \hline
    GPU Core Frequency & 1.30 GHz & LPDDR5 Packages & 8 \\ \hline
    SM Count & 16 & Dies / Package & 4 \\ \hline
    L2 Cache Capacity & 4 MB & Banks / Die & 16 (12 PIM, 4 Res.) \\ \hline
    LPDDR5 Bandwidth & 204.8 GB/s & Total PIM Units & 384 \\ \hline
    LPDDR5 Capacity & 64 GB & PIM Frequency & 200 MHz \\ \hline
    \multicolumn{4}{|c|}{\textbf{LPDDR5 Timing Parameters}} \\ \hline
    \multicolumn{2}{|c|}{$t_{RCD}=18.75$ ns, $t_{RP}=18.75$ ns} &
    \multicolumn{2}{c|}{$t_{RAS}=42.5$ ns, $t_{RFC}=280$ ns} \\ \hline
    \multicolumn{2}{|c|}{$t_{CCD\_L}=5$ ns, $t_{CCD\_S}=2.5$ ns} &
    \multicolumn{2}{c|}{$t_{RRD\_L}=5$ ns, $t_{FAW}=20$ ns} \\ \hline
  \end{tabular}%
  }
\end{table}
\subsection{Baselines}\label{sec:methodology-baselines}

\paragraph{\textbf{GPU Baselines}}

We evaluate the GPU paths on NVIDIA Jetson AGX Orin 64 GB in MAXN mode with fixed GPU and memory frequencies.
The frame-by-frame and multi-frame versions share the optimized CUDA backend.
AGX Orin + GSPIM Dataflow is a software-only GPU version implementing the complete multi-frame dataflow in CUDA.
It filters and generates Gaussians by window, reuses sorting results for the stable set, and renders frames with shared attribute batches sized to fit L2.
Full GSPIM retains this multi-frame flow and offloads selection and data reorganization to PPIM and HCF.
All paths share parameters, rendering settings, and compilation options. After warmup, CUDA events time stages and end-to-end execution, while Nsight Systems and Compute trace kernels and memory accesses.

\paragraph{\textbf{Specialized Accelerators}}

The specialized baselines are 3DGauCIM~\cite{3dgaucim}, QI-4DGS~\cite{thudgs}, and Relay-GS~\cite{relaygs}. 3DGauCIM uses FP16, 256 KB SRAM, 144 KB DCIM, grid and tile-block sizes of 4, a grouping threshold of 0.5, and eight AII-Sort partitions. QI-4DGS uses 16 fused recursive opacity units, eight tree-rendering PE lines, and 204 KB SRAM, selecting native generation or $2\times$/$4\times$ interpolation by motion criteria. Relay-GS reuses prior-frame sorting with periodic correction, a fine-grained pipeline, and parameter-sharing rasterization. We reconstruct cycle/power models from architecture/circuit data and reproduce original workloads and reference configurations, with maximum relative throughput/power errors below 1.5\%. All systems share the native explicit 4DGS model, camera path, resolution, precision, thresholds, and 256-bit LPDDR5-6400. 3DGauCIM, Relay-GS, and GSPIM retain native generation. QI-4DGS enables adaptive quadratic frame interpolation (AQFI) by default, with an AQFI-off control retaining opacity recurrence and tree rendering on the native generation path. ASIC area and power scale to 8 nm via DeepScaleTool interpolation to 10 nm and Samsung 10LPP-to-8LPP factors~\cite{sarangi2021deepscaletool,samsung2017eightlpp}. With external memory and shared control/interconnects counted, the remaining GSPIM system area budget determines replication counts from normalized tile areas including local storage. Tile organization, frequency, and replication stay fixed across scenes, and energy efficiency includes complete systems and external memory.

\begin{figure*}[!t]
  \centering
  \includegraphics[width=\textwidth]{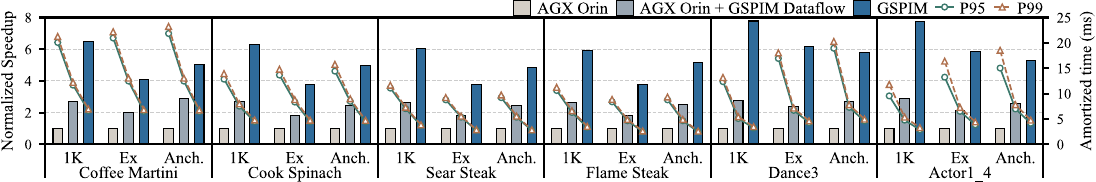}
  \caption{Steady-state throughput and amortized per-frame execution time (P95/P99) under identical ready-window submission.}
  \label{fig:overall-performance}
  \Description{Throughput normalized to AGX Orin on the left axis and P95 and P99 of amortized per-frame execution time in milliseconds on the right axis, for three execution paths, three workloads, and six scenes.}
\end{figure*}

\begin{figure*}[!t]
  \centering
  \includegraphics[width=\textwidth]{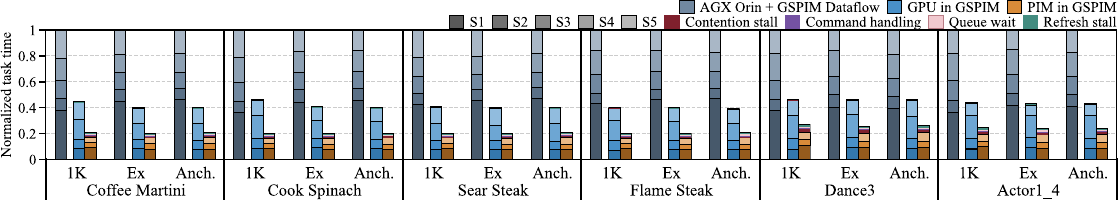}
  \caption{Stage-wise GPU and PIM task execution and waiting times.}
  \label{fig:stage-breakdown}
  \Description{Accumulated GPU and PIM task execution and waiting times normalized to AGX Orin plus GSPIM Dataflow, including contention, command handling, pre-execution queuing, and incremental refresh waits. Dark red denotes contention, purple denotes command handling, pale red denotes queuing, and green denotes refresh waiting.}
\end{figure*}

\begin{figure*}[!t]
  \centering
  \includegraphics[width=\textwidth]{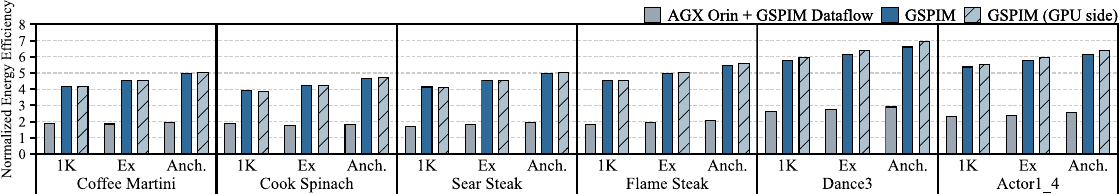}
  \caption{System and measured GPU task replay energy efficiency, normalized to matching frame-by-frame AGX Orin references.}
  \label{fig:energy-efficiency}
  \Description{System energy efficiency of AGX Orin plus GSPIM Dataflow and full GSPIM, alongside measured GPU-side task replay energy efficiency across six scenes and three models. System bars use the frame-by-frame AGX Orin system-energy reference. Hatched GPU-side bars use the matching measured AGX Orin GPU and SoC power-rail energy reference and exclude simulated PIM and LPDDR5 energy. Replay measurements cover execution and waiting.}
\end{figure*}
\subsection{Hardware Evaluation}\label{sec:hardware-evaluation}

\paragraph{\textbf{Cycle Simulation}}

We implement PPIM, HCF, and the DRAM-side command and status circuits in Chisel RTL. Their Verilator cycle models are integrated into Ramulator 2.0~\cite{ramulator2} for LPDDR5-6400 timing, including interface and cache-coherence overheads. The controller receives \texttt{PIM\_SELECT} and \texttt{PIM\_REORG} tasks and jointly schedules bank requests, GPU memory traces, and all-bank refresh under LPDDR5 timing, with refresh lasting 280 ns every 3.9~$\mu$s.

\paragraph{\textbf{Timing and Energy}}

Using Table~\ref{tab:platform}, we calibrate the timing model against AGX Orin bandwidth microbenchmarks and Nsight Compute memory traces. Under matching workloads, maximum relative errors are 1.4\% for bandwidth and 1.2\% for memory-access latency. We compare functional outputs with a per-primitive software implementation. We amortize window execution time per frame, accounting for task dependencies, measured GPU-stage times, simulated PIM time, refresh pauses, and GPU memory waits added by PIM traffic and refresh. GPU-side energy integrates onboard GPU and SoC rail measurements across replay execution and waiting. Replay follows the joint task schedule with functional-model inputs. PIM energy combines 1z switching-based dynamic energy and leakage over the full interval. LPDDR5 command and state traces determine read, write, refresh, and background energy. System and GPU-side efficiencies use matching frame-by-frame Orin references.

\paragraph{\textbf{Physical Implementation}}

We synthesize, place, and route added DRAM-side circuits using a 1z DRAM peripheral-logic library under baseline LPDDR5 bank- and die-periphery constraints. PPIM and L1 compactors sit beside PIM-bank row buffers, with HCF paths, command and completion circuits shared per die. PPIM operates at 200 MHz. Power uses cycle-simulation workload VCDs at 1.05 V, TT, and 25$^\circ$C. Icepak models the four-die stack, PCB, and module cooling in 3D with 40$^\circ$C inlet air at 1 m/s. LPDDR5, PIM, and SoC power traces include leakage updated with temperature.

\section{Evaluation}

\subsection{Performance and Quality}\label{sec:performance-quality}

\paragraph{\textbf{Performance}}

In Figure~\ref{fig:overall-performance}, window-level selection, stable-set sorting reuse, and attribute sharing give GPU dataflow 2.40$\times$ speedup and 46.8\%/48.2\% P95/P99 reductions over AGX Orin (geometric means). GPU selection and compaction require full-set scans and intermediate writes, prolonging preparation. Early PPIM selection and contiguous HCF packing reduce inactive-attribute transfers. Overlapping preparation with GPU generation and rendering reduces window waits. Full GSPIM achieves a 5.36$\times$ geometric-mean speedup over AGX Orin. Relative to GPU dataflow, mean amortized per-frame time, P95, and P99 fall by 55.2\%, 44.2\%, and 43.5\% (geometric means). Under strong nonrigid motion in Actor1\_4, Anchored 4DGS P99 falls from 7.81 to 4.77 ms.

Figure~\ref{fig:stage-breakdown} shows accumulated task execution and waiting times normalized to AGX Orin + GSPIM Dataflow.
GSPIM offloads S1, which occupies 36.1\% to 47.5\% of the GPU dataflow time, to PIM and reduces S3 and S5 GPU work through HCF reorganization, reducing summed GPU stage execution time by 54.1\% to 61.1\%.
PPIM selects next-window active records from local bank fields, alongside current-window GPU sorting and rendering that mainly access other banks.
Cross-frame L2 attribute reuse and HCF preparation of the next batch in another Batch Buffer reduce repeated reads and concurrent current-batch accesses.
Command handling and pre-execution queuing take 10.2 to 28.6~$\mu$s and 41.4 to 49.0~$\mu$s per die per window.
Contention stalls average 1.8\% and 7.7\% of total GPU and PIM task time.
PIM task execution totals 16.2\% to 21.1\% of GPU dataflow time and overlaps GPU execution.
Block and batch pipelines hide most refresh pauses, with throughput loss relative to disabled refresh averaging approximately 0.25\% and peaking at 0.35\%.

\paragraph{\textbf{Energy Efficiency}}

In Figure~\ref{fig:energy-efficiency}, software-only AGX Orin + GSPIM Dataflow achieves a 2.08$\times$ geometric-mean energy-efficiency improvement through sorting and attribute reuse.
CUDA selection and compaction limit further savings.
PPIM selection and HCF batching reduce GPU work and repeated transfers, yielding a 5.07$\times$ geometric-mean energy-efficiency improvement in measured GPU task replay.
Including PIM dynamic and leakage energy, together with LPDDR5 active and background energy, gives a 4.99$\times$ system improvement.
The AGX Orin software path could extend to desktop or server CUDA GPUs, with gains set by cross-frame reuse, cache capacity, and memory bandwidth.

\paragraph{\textbf{Comparison with Specialized Accelerators}}

\begin{figure}[!t]
  \centering
  \includegraphics[width=\columnwidth]{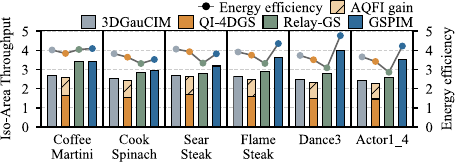}
  \caption{Iso-area throughput and energy efficiency.}
  \label{fig:accelerator-performance}
  \Description{Throughput and energy efficiency of 3DGauCIM, QI-4DGS, Relay-GS, and GSPIM across six dynamic scenes at equal total system area, normalized to AGX Orin. QI lower segments show AQFI-off throughput and upper segments show the AQFI-on increment. Energy-efficiency points refer to the full configurations, with AQFI enabled for QI.}
\end{figure}

In Figure~\ref{fig:accelerator-performance}, GSPIM reaches 3.43$\times$ geometric-mean native 4DGS throughput over AGX Orin at equal system area. 3DGauCIM~\cite{3dgaucim} combines coarse-grid selection, bucket and grouping reuse, and digital CIM. GSPIM applies record-level selection in programmable PIM before full attribute access, cutting traffic and improving throughput by 32.9\% and energy efficiency by 6.4\%. QI-4DGS~\cite{thudgs} reduces computation through interpolation, sorting merges, opacity recurrence, and tree compositing. GSPIM retains native generation and batches inputs across frames, exceeding AQFI-enabled QI throughput and energy efficiency by 39.8\% and 11.8\%. With AQFI off, QI has 1.57$\times$ geometric-mean throughput. GSPIM is 2.18$\times$ as fast on the same generation path. Relay-GS~\cite{relaygs} combines temporal sorting reuse, pipeline overlap, and rasterization parameter sharing. GSPIM batches generation inputs, attribute loads, and stable-set sorting across frames, amortizing transfers for 18.7\% higher throughput and 24.4\% higher energy efficiency. Temporal sparsity or large active sets amplify savings in Flame Steak, Dance3, and Actor1\_4, with Dance3 reaching 3.98$\times$. Greater per-frame generation and rasterization favor specialized arithmetic in Cook Spinach and Sear Steak. GSPIM retains at least 92.1\% of baseline energy efficiency.

\paragraph{\textbf{Quality}}

Table~\ref{tab:quality-gpu} reports scene-level quality relative to AGX Orin. GPU dataflow and full GSPIM differ from the baseline by at most 0.006 dB in PSNR and 0.0001 in SSIM at the reported precision. Conservative PPIM boundary handling adds no further difference at this precision. Table~\ref{tab:quality-accelerators} evaluates a separate native 4DGS model on the same six scenes, accounting for different Orin reference PSNRs from the three models in Table~\ref{tab:quality-gpu}. GSPIM differences peak at 0.010 dB and 0.0003, each below 0.04\%.

\begin{table}[t]
  \centering
  \caption{GSPIM rendering quality versus baseline.}
  \label{tab:quality-gpu}
  \resizebox{\columnwidth}{!}{
  \begin{tabular}{cc|cc}
    \toprule
    \textbf{Workload} & \textbf{Dataset} &
    \makecell{\textbf{PSNR}$\uparrow$\\\small\textbf{Base~/~Dataflow~/~GSPIM}} &
    \makecell{\textbf{SSIM}$\uparrow$\\\small\textbf{Base~/~Dataflow~/~GSPIM}} \\
    \midrule
    \multirow{6}{*}{4DGS-1K}
      & Coffee Martini & 27.890~/~27.889~/~27.889 & 0.9166~/~0.9166~/~0.9166 \\
      & Cook Spinach   & 32.947~/~32.945~/~32.945 & 0.9561~/~0.9560~/~0.9560 \\
      & Sear Steak     & 33.094~/~33.094~/~33.094 & 0.9627~/~0.9627~/~0.9627 \\
      & Flame Steak    & 33.271~/~33.268~/~33.268 & 0.9614~/~0.9614~/~0.9614 \\
      & Dance3         & 29.226~/~29.224~/~29.224 & 0.9063~/~0.9063~/~0.9063 \\
      & Actor1\_4      & 28.137~/~28.135~/~28.135 & 0.8449~/~0.8449~/~0.8449 \\
    \midrule
    \multirow{6}{*}{Ex4DGS}
      & Coffee Martini & 28.684~/~28.684~/~28.684 & 0.9184~/~0.9184~/~0.9184 \\
      & Cook Spinach   & 33.096~/~33.093~/~33.093 & 0.9563~/~0.9563~/~0.9563 \\
      & Sear Steak     & 33.435~/~33.434~/~33.434 & 0.9602~/~0.9602~/~0.9602 \\
      & Flame Steak    & 33.796~/~33.790~/~33.790 & 0.9628~/~0.9628~/~0.9628 \\
      & Dance3         & 29.310~/~29.309~/~29.309 & 0.9067~/~0.9067~/~0.9067 \\
      & Actor1\_4      & 28.153~/~28.153~/~28.153 & 0.8454~/~0.8454~/~0.8454 \\
    \midrule
    \multirow{6}{*}{\makecell{Anchored\\4DGS}}
      & Coffee Martini & 27.937~/~27.937~/~27.937 & 0.9047~/~0.9047~/~0.9047 \\
      & Cook Spinach   & 30.355~/~30.355~/~30.355 & 0.9387~/~0.9387~/~0.9387 \\
      & Sear Steak     & 32.559~/~32.559~/~32.559 & 0.9590~/~0.9590~/~0.9590 \\
      & Flame Steak    & 31.616~/~31.616~/~31.616 & 0.9489~/~0.9489~/~0.9489 \\
      & Dance3         & 28.195~/~28.195~/~28.195 & 0.8944~/~0.8944~/~0.8944 \\
      & Actor1\_4      & 27.334~/~27.334~/~27.334 & 0.8207~/~0.8207~/~0.8207 \\
    \bottomrule
  \end{tabular}}
\end{table}

\begin{table}[t]
  \centering
  \caption{Native 4DGS quality versus specialized accelerators.}
  \label{tab:quality-accelerators}
  \setlength{\tabcolsep}{2.5pt}
  \resizebox{\columnwidth}{!}{
  \begin{tabular}{l|ccccc||ccccc}
    \toprule
    \multirow{2}{*}{\textbf{Dataset}} &
    \multicolumn{5}{c||}{\textbf{PSNR}$\uparrow$} &
    \multicolumn{5}{c}{\textbf{SSIM}$\uparrow$} \\
    & \textbf{Orin} & \textbf{3DGauCIM} & \textbf{QI} & \textbf{Relay} & \textbf{GSPIM}
    & \textbf{Orin} & \textbf{3DGauCIM} & \textbf{QI} & \textbf{Relay} & \textbf{GSPIM} \\
    \midrule
    Sear Steak       & 32.920 & 32.780 & 32.730 & 32.867 & 32.917 & 0.9592 & 0.9568 & 0.9550 & 0.9567 & 0.9591 \\
    Coffee Martini   & 32.870 & 32.750 & 32.700 & 32.545 & 32.866 & 0.9335 & 0.9310 & 0.9290 & 0.9182 & 0.9333 \\
    Cook Spinach     & 31.770 & 31.660 & 31.610 & 31.212 & 31.764 & 0.9498 & 0.9474 & 0.9454 & 0.9345 & 0.9496 \\
    Flame Steak      & 33.130 & 32.950 & 32.900 & 31.983 & 33.124 & 0.9595 & 0.9566 & 0.9542 & 0.9542 & 0.9592 \\
    Dance3           & 29.180 & 28.980 & 28.930 & 28.654 & 29.171 & 0.9002 & 0.8960 & 0.8940 & 0.8928 & 0.8999 \\
    Actor1\_4        & 28.630 & 28.420 & 28.380 & 28.234 & 28.620 & 0.8733 & 0.8685 & 0.8670 & 0.8543 & 0.8730 \\
    \bottomrule
  \end{tabular}}
\end{table}

\begin{figure}[!t]
  \centering
  \includegraphics[width=\columnwidth]{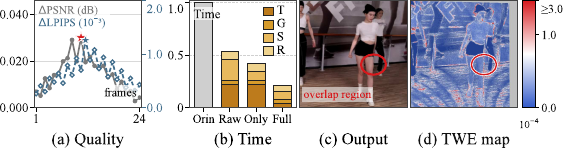}
  \caption{Sorting-reuse quality, execution time, and temporal error.}
  \label{fig:sorting-reuse-worst-case}
  \Description{Per-frame PSNR loss and LPIPS increase over 24 Dance3 frames, execution time normalized to Orin, a GSPIM-rendered frame, and the measured per-pixel motion-compensated temporal warping error map for frames 11 and 12.}
\end{figure}

\paragraph{\textbf{Worst-Case Error Analysis}}

In the 24-frame Dance3 camera-motion sequence with 4DGS-1K, nearby stable Gaussians can exchange depth order despite small depth changes, amplifying differences at high-contrast overlaps. Camera-space depth accounts for pose. Preserved attributes and per-frame projections confine errors to alpha-compositing order, weighted by opacity and foreground transmittance~\cite{3dgs}. In Figure~\ref{fig:sorting-reuse-worst-case}(a), PSNR loss peaks at 0.030 dB (frame 11) and LPIPS increase at approximately 0.0013. Consistent stable-set order avoids reordering jumps. Frames 11--12 yield mean motion-compensated temporal warping error (TWE) of approximately $1.25\times10^{-5}$ (Figure~\ref{fig:sorting-reuse-worst-case}(c,d)). Figure~\ref{fig:sorting-reuse-worst-case}(b) shows GPU window dataflow at 1.88$\times$ (full sorting) and 2.39$\times$ (reuse) over Orin, a 1.27$\times$ gain. Full GSPIM reaches 4.73$\times$ through near-memory selection, reorganization, and pipeline overlap.

\compactsubsection{Source of Performance Gain}\label{sec:source-of-performance-gain}

\paragraph{\textbf{Ablation Study}}

\begin{figure*}[!t]
  \centering
  \includegraphics[width=\textwidth]{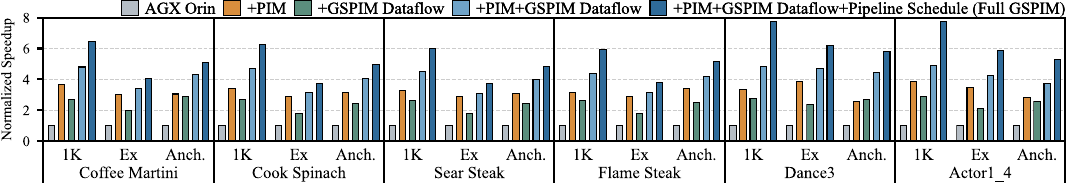}
  \caption{Ablation results across workloads and datasets.}
  \label{fig:ablation}
  \Description{Performance contributions of the PIM system, multi-frame dataflow, and hybrid pipeline scheduling.}
\end{figure*}

Figure~\ref{fig:ablation} isolates multi-frame dataflow, near-memory execution, and pipeline scheduling with five configurations. PIM alone preserves frame order while offloading activity decisions and compaction to PPIM and HCF. Combining dataflow and PIM starts each GPU stage after its PIM task, without cross-window or batch overlap. The keyframe intervals in Ex4DGS and anchor selection in Anchored 4DGS exclude more subsequent attributes with one decision. Enabling both designs yields 3.10$\times$--4.87$\times$ speedup. Full GSPIM overlaps near-memory preparation across windows and batches, adding 1.18$\times$--1.60$\times$ to reach 3.75$\times$--7.78$\times$ over the AGX Orin baseline.

\paragraph{\textbf{Polymorphic PIM Analysis}}

\begin{figure}[!t]
  \centering
  \includegraphics[width=\columnwidth]{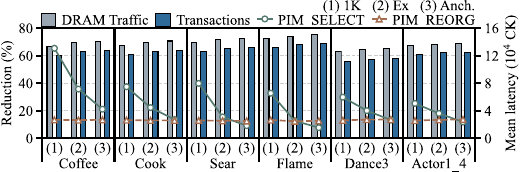}
  \caption{PIM traffic reductions and mean command latency.}
  \label{fig:pim-analysis}
  \Description{GPU-visible DRAM traffic and transaction reductions on the left axis, and mean die-level PIM_SELECT and PIM_REORG service latency in ten thousand DRAM CK cycles on the right axis.}
\end{figure}

Figure~\ref{fig:pim-analysis} shows GPU-visible DRAM traffic and transaction reductions from early PPIM selection and HCF compaction.
Across all workloads, these metrics decrease by 69.2\% and 62.7\% on average.
Modeled \texttt{PIM\_SELECT}/\texttt{PIM\_REORG} service latencies average $(1.54\text{--}13.07)\times10^4$/$(2.48\text{--}2.72)\times10^4$ DRAM cycles, including control and execution per die-level call.
Selection latency varies with programs and record counts.
Reorganization means include compaction and packing calls.
Ex4DGS and Anchored 4DGS exclude more associated attributes per decision.
Temporal sparsity benefits Flame Steak, while shorter windows and larger active sets limit gains in Dance3.

\paragraph{\textbf{Multi-Frame Rendering Dataflow Analysis}}

\begin{figure}[!t]
  \centering
  \includegraphics[width=\columnwidth]{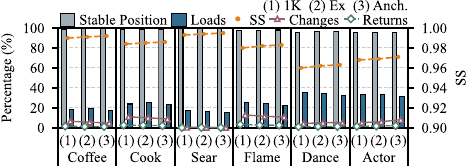}
  \caption{Dataflow stability, loads, and window adaptation.}
  \label{fig:dataflow-analysis}
  \Description{Depth stability, stable-set internal rank retention across all window frames, attribute loads, adjacent-window size changes, and immediate three-window returns across workloads and scenes.}
\end{figure}

In Figure~\ref{fig:dataflow-analysis}, stable-set internal rank retention across window frames reaches at least 95.8\% in Dance3 and 95.1\% in Actor1\_4. Threshold-boundary handling routes at most 0.0031\% of S3 Gaussians into per-frame sorting. Adjacent-window size changes and immediate three-window returns peak at 12.7\% and 3.18\%, with 13~$\mu$s/window control overhead in Actor1\_4. Motion-induced short windows raise attribute loads to 31.0\%--35.2\%, versus 15.4\%--16.9\% in Sear Steak. Cross-frame anchor sharing lowers mean Anchored 4DGS loads to 23.5\%.

\subsection{Scalability}\label{sec:scalability}

\paragraph{\textbf{Scalability to 3DGS}}

\begin{figure}[!t]
  \centering
  \includegraphics[width=\columnwidth]{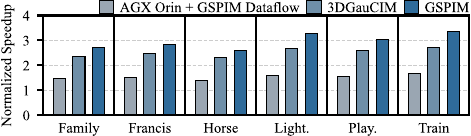}
  \caption{3DGS scalability relative to AGX Orin.}
  \label{fig:scalability-3dgs}
  \Description{Speedup of the GPU-only dataflow, 3DGauCIM, and GSPIM on static 3DGS scenes.}
\end{figure}

Figure~\ref{fig:scalability-3dgs} evaluates GSPIM on static 3DGS~\cite{3dgs}. Consecutive views share visible sets and depth orders, enabling reuse of visibility selection, sorting, and attribute loading. Across six Tanks and Temples scenes~\cite{tanksandtemples}, the GPU dataflow and full GSPIM achieve geometric-mean speedups of 1.54$\times$ and 2.96$\times$, respectively. Full GSPIM improves over the 2.52$\times$ of 3DGauCIM~\cite{3dgaucim} by 1.17$\times$ because visibility-driven compaction excludes primitives retained by coarse grids and reuses their layout and depth order.

\paragraph{\textbf{Scalability Across GPU Platforms}}

\begin{figure}[!t]
  \centering
  \includegraphics[width=\columnwidth]{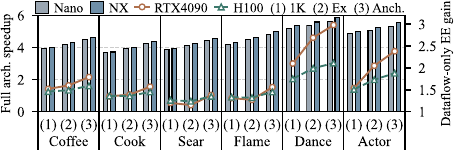}
  \caption{Scalability across GPU platforms.}
  \label{fig:scalability-gpu}
  \Description{Full GSPIM speedups on Jetson Orin Nano and NX use the left axis. Software-only GSPIM Dataflow energy-efficiency gains on RTX4090 and H100 SXM use the right axis. Each GPU is normalized to its own optimized frame-by-frame baseline across three 4DGS methods and six dynamic scenes.}
\end{figure}

Figure~\ref{fig:scalability-gpu} reports full GSPIM speedups of 4.52/4.63$\times$ on Nano/NX and Dataflow-only energy-efficiency gains of 1.65/1.52$\times$ on RTX4090/H100 (geometric means over optimized per-GPU baselines). Small Nano/NX caches increase baseline reloads, favoring PPIM filtering/HCF compaction. Sharing L2-sized batches and stable-set orders within a window reduces memory traffic and sorting energy. H100 bandwidth reduces baseline stalls and relative energy savings from attribute reuse.

\paragraph{\textbf{Scene Scale and Resolution}}

\begin{figure}[!t]
  \centering
  \includegraphics[width=\columnwidth]{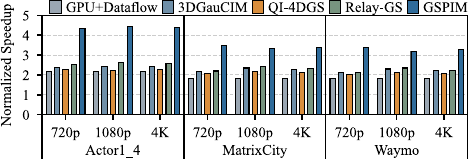}
  \caption{Scene scale and resolution.}
  \label{fig:scalability-scale-resolution}
  \Description{Speedups of GPU+Dataflow, 3DGauCIM, QI-4DGS, Relay-GS, and GSPIM for Actor1\_4, MatrixCity, and Waymo, normalized to the matching scene and resolution on AGX Orin. All values are author-confirmed measurements.}
\end{figure}

% All displayed scene-resolution speedups are author-confirmed measurements.
Figure~\ref{fig:scalability-scale-resolution} compares Actor1\_4, MatrixCity~\cite{matrixcity} (23.7M Gaussians), and Waymo~\cite{waymo} at 720p, 1080p, and 4K with fixed hardware. GPU+Dataflow and GSPIM reach geometric-mean speedups of 1.93$\times$ and 3.66$\times$ over Orin. GSPIM exceeds 3DGauCIM, QI-4DGS, and Relay-GS by 1.59$\times$, 1.71$\times$, and 1.55$\times$. PPIM selection and HCF compaction avoid GPU transfers of inactive data. Depth-based classification and window adjustment adapt sorting reuse to camera and object motion while sorting unstable primitives per frame. L2-sized batches retain attribute reuse as intermediate data grows with scale and resolution. At 4K, attribute reuse reduces repeated loads while HCF prepares the next batch during pixel compositing.

\subsection{Sensitivity Study}\label{sec:sensitivity-study}

\paragraph{\textbf{Stability Threshold}}

\begin{figure}[!t]
  \centering
  \includegraphics[width=\columnwidth]{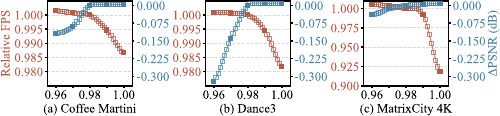}
  \caption{Sensitivity to the stability threshold $\tau_s$.}
  \label{fig:sensitivity-tau-s}
  \Description{Relative FPS and PSNR changes with the stability threshold in Coffee Martini, Dance3, and MatrixCity at 4K, referenced to the 0.98 setting. Filled squares mark input data points and hollow squares show interpolation.}
\end{figure}

Figure~\ref{fig:sensitivity-tau-s} sweeps $\tau_s$ around 0.98. Equation~\ref{eq:stability-score} bounds mean normalized depth variation for stable primitives by $-\ln\tau_s$, or 0.0202 at $\tau_s=0.98$. Common positive affine transforms of camera depths and clipping bounds preserve $d_i$ and classification across depth units. The local criterion omits Gaussian count and image resolution. Lower thresholds admit more depth variation, while higher thresholds reduce throughput for at most 0.01 dB gain. Across all cases, 0.98 stays within 0.6\% of peak FPS and 0.012 dB of full-sorting PSNR.

\paragraph{\textbf{Window Adjustment Threshold}}

\begin{figure}[!t]
  \centering
  \includegraphics[width=\columnwidth]{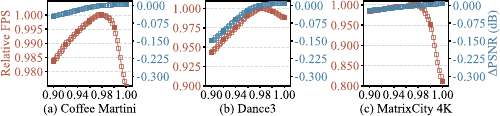}
  \caption{Sensitivity to the window-adjustment threshold $\tau_w$.}
  \label{fig:sensitivity-tau-w}
  \Description{Relative FPS and PSNR changes with the window-adjustment threshold in Coffee Martini, Dance3, and MatrixCity at 4K, referenced to the 0.98 setting. Filled squares mark input data points and hollow squares show interpolation.}
\end{figure}

Figure~\ref{fig:sensitivity-tau-w} sweeps $\tau_w$ with the same 0.98 reference. Replicating the stability-score distribution leaves its mean unchanged, tying window adjustment to depth dynamics rather than primitive count. One-frame changes within the two-to-five-frame range bound the control response. Lower thresholds allow longer windows with greater variation, while higher thresholds shorten cross-frame reuse. Motion changes adjustment frequency, but all groups preserve the same unimodal trend. Across all representations and datasets, 0.98 remains within 0.8\% of peak FPS and 0.012 dB of full-sorting PSNR.

\paragraph{\textbf{Reserved Workspace Capacity}}

\begin{figure}[!t]
  \centering
  \includegraphics[width=\columnwidth]{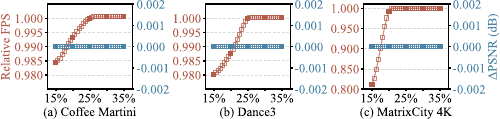}
  \caption{Sensitivity to reserved workspace capacity.}
  \label{fig:sensitivity-active-buffer}
  \Description{Relative FPS and PSNR changes with the fraction of each die allocated to HCF-managed buffers in Coffee Martini, Dance3, and MatrixCity at 4K, referenced to the 25 percent allocation. Filled squares mark input data points and hollow squares show interpolation.}
\end{figure}

Figure~\ref{fig:sensitivity-active-buffer} varies the HCF workspace for the Active Buffer, classification sequences, and Batch Buffers. Live-range sharing sizes capacity for peak occupancy, and L2-sized batches bound Batch Buffer storage. Active Buffer and classification storage scale with the active set. Smaller allocations trigger GPU reorganization, reducing preparation-rendering overlap. Throughput saturates at 25\% (four Res. Banks), within 0.1\% of larger capacities across all three scenes. Reorganization preserves attributes, record associations, render order, and PSNR.

\subsection{Hardware Overhead}\label{sec:hardware-overhead}

Across 32 dies, Table~\ref{tab:area-power} reports 15.9096 mm$^2$ added cell area, 89.9\% in bank-level circuits, chiefly PPIM and L1 compactors replicated in 384 PIM banks. Including routing, power distribution, and spacing, the 18.7172 mm$^2$ physical footprint is 0.5849 mm$^2$ per die, or 0.96\% of a 60.92 mm$^2$ 16 Gb LPDDR5 reference die~\cite{he2025lpspec}. The masked column-selection logic adds 0.28 ns within column-access timing margins, preserving external read/write latencies and column-command intervals. Added active power is 3.8363 W (3.2582 W dynamic, 0.5781 W leakage), or 0.120 W per die and 0.480 W per four-die package, with 50.6\% in the fixed-point datapath and L1 compactor. Thermal analysis gives an approximately 70$^\circ$C steady-state peak package case temperature, about 15$^\circ$C below the 85$^\circ$C WT-grade operating limit~\cite{micron2021lpddr5}.

\begin{table}[t]
  \centering
  \caption{Area and power breakdown of GSPIM additions to LPDDR5.}
  \label{tab:area-power}
  \scriptsize
  \setlength{\tabcolsep}{0.8pt}
  \renewcommand{\arraystretch}{0.92}
  % Values confirmed from target 1z synthesis/physical-design reports on 2026-10-03.
  % Confirmation records user-supplied evidence. Raw reports have not been inspected.
  % Area is cell area. Power combines dynamic power and leakage.
  % All entries are aggregated over the instance counts in the group labels.
  \begin{tabular*}{\columnwidth}{@{}l@{\hspace{3pt}}r@{\hspace{5pt}}r@{\extracolsep{\fill}}l@{\hspace{3pt}}r@{\hspace{5pt}}r@{\hspace{2pt}}}
    \toprule
    \multirow{2}{*}{\textbf{Component}} & \multicolumn{1}{c}{\textbf{Area}} & \multicolumn{1}{c}{\textbf{Power}} & \multirow{2}{*}{\textbf{Component}} & \multicolumn{1}{c}{\textbf{Area}} & \multicolumn{1}{c}{\textbf{Power}} \\
    \multicolumn{1}{c}{} & \multicolumn{1}{c}{(mm$^2$)} & \multicolumn{1}{c}{(mW)} & \multicolumn{1}{c}{} & \multicolumn{1}{c}{(mm$^2$)} & \multicolumn{1}{c}{(mW)} \\
    \midrule
    \multicolumn{6}{@{}l}{\textbf{Bank Logic, $\times384$ (14.3064 mm$^2$, 3,169.1 mW)}} \\
    \quad Program Buf. & 7.2762 & 313.9 & \quad Instr. Seq. & 0.3073 & 159.8 \\
    \quad Fixed Datapath & 3.5143 & 1,437.9 & \quad Working RF & 1.3980 & 194.6 \\
    \quad Reduction Unit & 0.3132 & 169.8 & \quad Selection Unit & 0.1724 & 87.3 \\
    \quad L1 Compactor & 0.8011 & 504.1 & \quad Mask Col. Path & 0.1053 & 52.1 \\
    \quad Bank Interface & 0.1657 & 102.9 & \quad FP32 Converter & 0.2529 & 146.7 \\
    \midrule
    \multicolumn{6}{@{}l}{\textbf{Die Logic, $\times32$ (1.6028 mm$^2$, 665.8 mW)}} \\
    \quad Layout Regs & 0.0414 & 2.6 & \quad Map Scanner & 0.6156 & 52.9 \\
    \quad Dest. Allocator & 0.0224 & 12.6 & \quad Addr. Gen. & 0.0282 & 14.5 \\
    \quad Read Coalescer & 0.1780 & 51.1 & \quad Data Router & 0.0783 & 47.6 \\
    \quad Write Packer & 0.2021 & 62.4 & \quad Task Scheduler & 0.0256 & 13.3 \\
    \quad Task RF & 0.1588 & 26.0 & \quad Access Arbiter & 0.0173 & 8.6 \\
    \quad Cmd. Frontend & 0.0668 & 129.9 & \quad Completion Unit & 0.0414 & 23.3 \\
    \quad Bank Ctrl & 0.0711 & 36.8 & \quad Clock Ctrl & 0.0558 & 184.2 \\
    \midrule
    Clock Div. ($\times32$) & 0.0004 & 1.4 & \textbf{Total} & \textbf{15.9096} & \textbf{3,836.3} \\
    \bottomrule
  \end{tabular*}%
\end{table}

\section{Related Work}

\paragraph{\textbf{3DGS Hardware Acceleration}}

3DGS hardware spans dedicated and GPU-attached rendering~\cite{gscore,gbu,gstg,ls-gaussian,gaupre,gcc,pei2026degs,flicker,chang129fps}, cross-view reuse~\cite{lumina,neo}, NPU execution~\cite{orange}, ray-based and large-scale rendering~\cite{grtx,nebula}, and training or SLAM~\cite{gsarch,gauspu,react3d,localgs,boa3dgs,cambricongs,splatonic,ags}. These accelerate generated static Gaussians or online reconstruction. Neo, closest in depth reuse, updates fixed-scene order across views. GSPIM filters dynamic records before generation and compacts attributes for multi-frame execution.

\paragraph{\textbf{4DGS Hardware Acceleration}}

3DGauCIM~\cite{3dgaucim} combines coarse spatiotemporal indexing, preceding-frame reuse, and digital CIM rendering. QI-4DGS~\cite{thudgs} generates intermediate-frame 2DGS through adaptive quadratic interpolation, accelerating opacity computation and rendering. Relay-GS~\cite{relaygs} selectively reuses preceding-frame sorting with periodic correction. Specter~\cite{specter} unifies modeling and rendering, skips modeling undeformed tiles, and renders by contribution. Relay-GS is closest in depth reuse but specializes sorting for one rendering path. GSPIM retains native GPU generation across representations, combining configurable temporal selection, near-memory compaction, and window-level depth reuse.

\section{Conclusion}

GSPIM accelerates edge 4DGS with representation-specific PPIM selection and HCF compaction of active attributes before GPU loads. Cross-frame attribute and stable-set sorting reuse reduces repeated work within L2-sized batches, while hybrid scheduling overlaps near-memory preparation and GPU rendering. Across three 4DGS methods and six dynamic scenes, speedup and energy-efficiency gains over AGX Orin reach 5.36$\times$ and 4.99$\times$ (geometric means), with scene-level PSNR differences at most 0.006 dB. Across 32 1z DRAM dies, the added physical footprint and active power total 18.7172~mm$^2$ and 3.8363~W.

%%%%%%% -- PAPER CONTENT ENDS -- %%%%%%%%

%%%%%%%%% -- BIB STYLE AND FILE -- %%%%%%%%
\bibliographystyle{IEEEtranS}
\bibliography{paper_refs}

@String{Computing = "Computing" }

@String{Computer = "{IEEE} Computer" }

@article{3dgaucim,
  author     = {Huang, Wei-Hsing and Shih, Cheng-Jhih and Su, Jian-Wei and Wang, Samuel and Garg, Vaidehi and Kong, Yuyao and Tien, Jen-Chun and Li, Nealson and Raychowdhury, Arijit and Chang, Meng-Fan and Lin, Yingyan (Celine) and Yu, Shimeng},
  title      = {{3DGauCIM}: Accelerating Static/Dynamic {3D} Gaussian Splatting via Digital {CIM} for High Frame Rate Real-Time Edge Rendering},
  year       = {2026},
  issue_date = {July 2026},
  publisher  = {Association for Computing Machinery},
  address    = {New York, NY, USA},
  volume     = {31},
  number     = {4},
  issn       = {1084-4309},
  url        = {https://doi.org/10.1145/3793555},
  doi        = {10.1145/3793555},
  journal    = {ACM Trans. Des. Autom. Electron. Syst.},
  month      = mar,
  articleno  = {81},
  numpages   = {18},
  pages        = {81:1--81:18}
}

@article{3dgs,
  author    = {Bernhard Kerbl and
               Georgios Kopanas and
               Thomas Leimk{\"{u}}hler and
               George Drettakis},
  title     = {{3D} {Gaussian} Splatting for Real-Time Radiance Field Rendering},
  journal   = {{ACM} Trans. Graph.},
  volume    = {42},
  number    = {4},
  pages     = {139:1--139:14},
  year      = {2023},
  url       = {https://doi.org/10.1145/3592433},
  doi       = {10.1145/3592433},
  bibsource = {dblp computer science bibliography, https://dblp.org}
}

@inproceedings{4dgs1k,
  title     = {1000+ {FPS} {4D} {Gaussian} Splatting for Dynamic Scene Rendering},
  author    = {Yuheng Yuan and Qiuhong Shen and Xingyi Yang and Xinchao Wang},
  booktitle = {Advances in Neural Information Processing Systems},
  year      = {2025},
  url       = {https://proceedings.neurips.cc/paper_files/paper/2025/hash/6c39d1a7eecfd98570d74bf7efec1be7-Abstract-Conference.html},
  volume       = {38},
  pages        = {83135--83153},
  doi          = {10.52202/085713-2507}
}

@inproceedings{ex4dgs,
  author    = {Junoh Lee and
               Changyeon Won and
               Hyunjun Jung and
               Inhwan Bae and
               Hae-Gon Jeon},
  title     = {Fully Explicit Dynamic {Gaussian} Splatting},
  booktitle = {Advances in Neural Information Processing Systems},
  volume    = {37},
  year      = {2024},
  url       = {https://proceedings.neurips.cc/paper_files/paper/2024/hash/09b47a77997b7dd7d2b26bd8ff769392-Abstract-Conference.html},
  doi       = {10.52202/079017-0174},
  pages        = {5384--5409}
}

@inproceedings{4drotorgs,
  author    = {Yuanxing Duan and
               Fangyin Wei and
               Qiyu Dai and
               Yuhang He and
               Wenzheng Chen and
               Baoquan Chen},
  editor    = {Andres Burbano and
               Denis Zorin and
               Wojciech Jarosz},
  title     = {{4D}-Rotor {Gaussian} Splatting: Towards Efficient Novel View Synthesis
for Dynamic Scenes},
  booktitle = {{ACM} {SIGGRAPH} 2024 Conference Papers, {SIGGRAPH} 2024, Denver,
               CO, USA, 27 July 2024- 1 August 2024},
  pages     = {87:1--87:11},
  publisher = {{ACM}},
  year      = {2024},
  url       = {https://doi.org/10.1145/3641519.3657463},
  doi       = {10.1145/3641519.3657463},
  bibsource = {dblp computer science bibliography, https://dblp.org}
}

@inproceedings{anchored4dgs,
  author    = {Yilong Li and
               Bo Pang and
               Yisong Chen and
               Guoping Wang},
  editor    = {Taku Komura and
               Michael Wimmer and
               Hongbo Fu},
  title     = {Anchored {4D} {Gaussian} Splatting for Dynamic Novel View Synthesis},
  booktitle = {Proceedings of the {SIGGRAPH} Asia 2025 Conference Papers, {SA} Conference
               Papers 2025, Hong Kong, December 15-18, 2025},
  pages     = {73:1--73:11},
  publisher = {{ACM}},
  year      = {2025},
  url       = {https://doi.org/10.1145/3757377.3763898},
  doi       = {10.1145/3757377.3763898},
  bibsource = {dblp computer science bibliography, https://dblp.org}
}

@inproceedings{aquabolt-xl,
  author    = {Jin Hyun Kim and
               Shinhaeng Kang and
               Sukhan Lee and
               Hyeonsu Kim and
               Woongjae Song and
               Yuhwan Ro and
               Seungwon Lee and
               David Wang and
               Hyunsung Shin and
               BengSeng Phuah and
               Jihyun Choi and
               Jinin So and
               YeonGon Cho and
               Joon{-}Ho Song and
               Jangseok Choi and
               Jeonghyeon Cho and
               Kyomin Sohn and
               Young{-}Soo Sohn and
               Kwang{-}Il Park and
               Nam Sung Kim},
  title     = {{Aquabolt-XL}: Samsung {HBM2-PIM} with in-memory processing for {ML}
accelerators and beyond},
  booktitle = {{IEEE} Hot Chips 33 Symposium, {HCS} 2021, Palo Alto, CA, USA, August
               22-24, 2021},
  pages     = {1--26},
  publisher = {{IEEE}},
  year      = {2021},
  url       = {https://doi.org/10.1109/HCS52781.2021.9567191},
  doi       = {10.1109/HCS52781.2021.9567191},
  bibsource = {dblp computer science bibliography, https://dblp.org}
}

@inproceedings{centralizedpim,
  author    = {Yufeng Gu and
               Alireza Khadem and
               Sumanth Umesh and
               Ning Liang and
               Xavier Servot and
               Onur Mutlu and
               Ravi R. Iyer and
               Reetuparna Das},
  editor    = {Lieven Eeckhout and
               Georgios Smaragdakis and
               Katai Liang and
               Adrian Sampson and
               Martha A. Kim and
               Christopher J. Rossbach},
  title     = {{PIM} Is All You Need: {A} {CXL}-Enabled {GPU}-Free System for Large Language
Model Inference},
  booktitle = {Proceedings of the 30th {ACM} International Conference on Architectural
               Support for Programming Languages and Operating Systems, Volume 2,
               {ASPLOS} 2025, Rotterdam, Netherlands, 30 March 2025 - 3 April 2025},
  pages     = {862--881},
  publisher = {{ACM}},
  year      = {2025},
  url       = {https://doi.org/10.1145/3676641.3716267},
  doi       = {10.1145/3676641.3716267},
  bibsource = {dblp computer science bibliography, https://dblp.org}
}

@inproceedings{desiregs,
  author    = {Chensheng Peng and
               Chengwei Zhang and
               Yixiao Wang and
               Chenfeng Xu and
               Yichen Xie and
               Wenzhao Zheng and
               Kurt Keutzer and
               Masayoshi Tomizuka and
               Wei Zhan},
  title     = {{DeSiRe-GS}: {4D} Street {Gaussians} for Static-Dynamic Decomposition and
Surface Reconstruction for Urban Driving Scenes},
  booktitle = {{IEEE/CVF} Conference on Computer Vision and Pattern Recognition,
               {CVPR} 2025, Nashville, TN, USA, June 11-15, 2025},
  pages     = {6782--6791},
  publisher = {Computer Vision Foundation / {IEEE}},
  year      = {2025},
  url       = {https://openaccess.thecvf.com/content/CVPR2025/html/Peng_DeSiRe-GS_4D_Street_Gaussians_for_Static-Dynamic_Decomposition_and_Surface_Reconstruction_CVPR_2025_paper.html},
  doi       = {10.1109/CVPR52734.2025.00636},
  bibsource = {dblp computer science bibliography, https://dblp.org}
}

@inproceedings{easyvolcap,
  author    = {Zhen Xu and
               Tao Xie and
               Sida Peng and
               Haotong Lin and
               Qing Shuai and
               Zhiyuan Yu and
               Guangzhao He and
               Jiaming Sun and
               Hujun Bao and
               Xiaowei Zhou},
  editor    = {June Kim and
               Rajesh Sharma},
  title     = {{EasyVolcap}: Accelerating Neural Volumetric Video Research},
  booktitle = {{SIGGRAPH} Asia 2023 Technical Communications, {SA} Technical Communications
               2023, Sydney, NSW, Australia, December 12-15, 2023},
  pages     = {22:1--22:4},
  publisher = {{ACM}},
  year      = {2023},
  url       = {https://doi.org/10.1145/3610543.3626173},
  doi       = {10.1145/3610543.3626173},
  bibsource = {dblp computer science bibliography, https://dblp.org}
}

@inproceedings{flashgs,
  author    = {Guofeng Feng and
               Siyan Chen and
               Rong Fu and
               Zimu Liao and
               Yi Wang and
               Tao Liu and
               Boni Hu and
               Linning Xu and
               Zhilin Pei and
               Hengjie Li and
               Xiuhong Li and
               Ninghui Sun and
               Xingcheng Zhang and
               Bo Dai},
  title     = {{FlashGS}: Efficient {3D} {Gaussian} Splatting for Large-scale and High-resolution
Rendering},
  booktitle = {{IEEE/CVF} Conference on Computer Vision and Pattern Recognition,
               {CVPR} 2025, Nashville, TN, USA, June 11-15, 2025},
  pages     = {26652--26662},
  publisher = {Computer Vision Foundation / {IEEE}},
  year      = {2025},
  url       = {https://openaccess.thecvf.com/content/CVPR2025/html/Feng_FlashGS_Efficient_3D_Gaussian_Splatting_for_Large-scale_and_High-resolution_Rendering_CVPR_2025_paper.html},
  doi       = {10.1109/CVPR52734.2025.02482},
  bibsource = {dblp computer science bibliography, https://dblp.org}
}

@article{gsplat,
  author  = {Vickie Ye and
             Ruilong Li and
             Justin Kerr and
             Matias Turkulainen and
             Brent Yi and
             Zhuoyang Pan and
             Otto Seiskari and
             Jianbo Ye and
             Jeffrey Hu and
             Matthew Tancik and
             Angjoo Kanazawa},
  title   = {{gsplat}: An Open-Source Library for {Gaussian} Splatting},
  journal = {Journal of Machine Learning Research},
  volume  = {26},
  number  = {34},
  pages   = {1--17},
  year    = {2025},
  url          = {https://jmlr.org/papers/v26/24-1476.html}
}

@inproceedings{gazevis,
  author    = {Yonas T. Tefera and
               Dario Mazzanti and
               Sara Anastasi and
               Darwin G. Caldwell and
               Paolo Fiorini and
               Nikhil Deshpande},
  title     = {Towards Gaze-contingent Visualization of Real-time {3D} Reconstructed
Remote Scenes in Mixed Reality},
  booktitle = {Proceedings of the 2023 6th International Conference on Advances in
               Robotics, {AIR} 2023, Ropar, India, July 5-8, 2023},
  pages     = {21:1--21:6},
  publisher = {{ACM}},
  year      = {2023},
  url       = {https://doi.org/10.1145/3610419.3610440},
  doi       = {10.1145/3610419.3610440},
  bibsource = {dblp computer science bibliography, https://dblp.org}
}

@inproceedings{gcc,
  author    = {Minnan Pei and
               Gang Li and
               Junwen Si and
               Zeyu Zhu and
               Zitao Mo and
               Peisong Wang and
               Zhuoran Song and
               Xiaoyao Liang and
               Jian Cheng},
  title     = {{GCC:} {A} {3DGS} Inference Architecture with {Gaussian}-Wise and Cross-Stage
Conditional Processing},
  booktitle = {Proceedings of the 58th {IEEE/ACM} International Symposium on Microarchitecture,
               {MICRO} 2025, Seoul, Republic of Korea, October 18-22, 2025},
  pages     = {1824--1837},
  publisher = {{ACM}},
  year      = {2025},
  url       = {https://doi.org/10.1145/3725843.3756072},
  doi       = {10.1145/3725843.3756072},
  bibsource = {dblp computer science bibliography, https://dblp.org}
}

@inproceedings{gbu,
  author    = {Zhifan Ye and
Yonggan Fu and
Jingqun Zhang and
Leshu Li and
Yongan Zhang and
Sixu Li and
Cheng Wan and
Chenxi Wan and
Chaojian Li and
Sreemanth Prathipati and
Yingyan Celine Lin},
  title     = {{Gaussian Blending Unit}: An Edge {GPU} Plug-in for Real-Time {Gaussian}-Based Rendering in {AR/VR}},
  booktitle = {Proceedings of the 31st {IEEE} International Symposium on High Performance Computer Architecture},
  pages     = {353--365},
  publisher = {{IEEE}},
  year      = {2025},
  doi       = {10.1109/HPCA61900.2025.00036}
}

@inproceedings{hexplane,
  author    = {Ang Cao and
               Justin Johnson},
  title     = {{HexPlane}: {A} Fast Representation for Dynamic Scenes},
  booktitle = {{IEEE/CVF} Conference on Computer Vision and Pattern Recognition,
               {CVPR} 2023, Vancouver, BC, Canada, June 17-24, 2023},
  pages     = {130--141},
  publisher = {{IEEE}},
  year      = {2023},
  url       = {https://doi.org/10.1109/CVPR52729.2023.00021},
  doi       = {10.1109/CVPR52729.2023.00021},
  bibsource = {dblp computer science bibliography, https://dblp.org}
}

@inproceedings{hybridradixsort,
  author    = {Elias Stehle and
               Hans{-}Arno Jacobsen},
  editor    = {Semih Salihoglu and
               Wenchao Zhou and
               Rada Chirkova and
               Jun Yang and
               Dan Suciu},
  title     = {A Memory Bandwidth-Efficient Hybrid Radix Sort on {GPUs}},
  booktitle = {Proceedings of the 2017 {ACM} International Conference on Management
               of Data, {SIGMOD} Conference 2017, Chicago, IL, USA, May 14-19, 2017},
  pages     = {417--432},
  publisher = {{ACM}},
  year      = {2017},
  url       = {https://doi.org/10.1145/3035918.3064043},
  doi       = {10.1145/3035918.3064043},
  bibsource = {dblp computer science bibliography, https://dblp.org}
}

@inproceedings{imprint,
  author    = {Ersin Cukurtas and
               Kavish Ranawella and
               Kevin Skadron and
               Mircea Stan},
  title     = {{IMPRINT:} In-Memory Processing with Indirect Addressing Techniques
for {GPU}-hosted {HBM-PIM}},
  booktitle = {Proceedings of the International Symposium on Memory Systems, MemSys
               2025, Washington, USA, October 6-8, 2025},
  pages     = {165--176},
  publisher = {{ACM}},
  year      = {2025},
  url       = {https://doi.org/10.1145/3767110.3767121},
  doi       = {10.1145/3767110.3767121},
  bibsource = {dblp computer science bibliography, https://dblp.org}
}

@inproceedings{kplanes,
  author    = {Sara Fridovich{-}Keil and
               Giacomo Meanti and
               Frederik Rahb{\ae}k Warburg and
               Benjamin Recht and
               Angjoo Kanazawa},
  title     = {{K-Planes}: Explicit Radiance Fields in Space, Time, and Appearance},
  booktitle = {{IEEE/CVF} Conference on Computer Vision and Pattern Recognition,
               {CVPR} 2023, Vancouver, BC, Canada, June 17-24, 2023},
  pages     = {12479--12488},
  publisher = {{IEEE}},
  year      = {2023},
  url       = {https://doi.org/10.1109/CVPR52729.2023.01201},
  doi       = {10.1109/CVPR52729.2023.01201},
  bibsource = {dblp computer science bibliography, https://dblp.org}
}

@inproceedings{ls-gaussian,
  author    = {Wei, Linye and Tang, Jiajun and Fei, Fan and Shi, Boxin and Wang, Runsheng and Li, Meng},
  booktitle = {2025 IEEE/ACM International Conference On Computer Aided Design (ICCAD)},
  title     = {No Redundancy, No Stall: Lightweight Streaming {3D} {Gaussian} Splatting for Real-time Rendering},
  year      = {2025},
  volume    = {},
  number    = {},
  pages     = {1-9},
  doi       = {10.1109/ICCAD66269.2025.11240950}
}

@inproceedings{mega,
  author    = {Zhang, Xinjie and Liu, Zhening and Zhang, Yifan and Ge, Xingtong and He, Dailan and Xu, Tongda and Wang, Yan and Lin, Zehong and Yan, Shuicheng and Zhang, Jun},
  title     = {{MEGA}: Memory-Efficient {4D} {Gaussian} Splatting for Dynamic Scenes},
  booktitle = {Proceedings of the IEEE/CVF International Conference on Computer Vision (ICCV)},
  month     = {October},
  year      = {2025},
  pages     = {27828-27838},
  url          = {https://openaccess.thecvf.com/content/ICCV2025/html/Zhang_MEGA_Memory-Efficient_4D_Gaussian_Splatting_for_Dynamic_Scenes_ICCV_2025_paper.html}
}

@inproceedings{mergepath,
  author    = {Saher Odeh and
               Oded Green and
               Zahi Mwassi and
               Oz Shmueli and
               Yitzhak Birk},
  title     = {{Merge Path} - Parallel Merging Made Simple},
  booktitle = {26th {IEEE} International Parallel and Distributed Processing Symposium
               Workshops {\&} PhD Forum, {IPDPS} 2012, Shanghai, China, May 21-25,
               2012},
  pages     = {1611--1618},
  publisher = {{IEEE} Computer Society},
  year      = {2012},
  url       = {https://doi.org/10.1109/IPDPSW.2012.202},
  doi       = {10.1109/IPDPSW.2012.202},
  bibsource = {dblp computer science bibliography, https://dblp.org}
}

@inproceedings{n3v,
  author    = {Tianye Li and
               Mira Slavcheva and
               Michael Zollh{\"{o}}fer and
               Simon Green and
               Christoph Lassner and
               Changil Kim and
               Tanner Schmidt and
               Steven Lovegrove and
               Michael Goesele and
               Richard A. Newcombe and
               Zhaoyang Lv},
  title     = {Neural {3D} Video Synthesis from Multi-view Video},
  booktitle = {{IEEE/CVF} Conference on Computer Vision and Pattern Recognition,
               {CVPR} 2022, New Orleans, LA, USA, June 18-24, 2022},
  pages     = {5511--5521},
  publisher = {{IEEE}},
  year      = {2022},
  url       = {https://doi.org/10.1109/CVPR52688.2022.00544},
  doi       = {10.1109/CVPR52688.2022.00544},
  bibsource = {dblp computer science bibliography, https://dblp.org}
}

@inproceedings{neo,
  author    = {Oh, Changhun and Oh, Seongryong and Hwang, Jinwoo and Kim, Yoonsung and Sharma, Hardik and Park, Jongse},
  title     = {{Neo}: Real-Time On-Device {3D} {Gaussian} Splatting with Reuse-and-Update Sorting Acceleration},
  booktitle = {Proceedings of the 31st ACM International Conference on Architectural Support for Programming Languages and Operating Systems, Volume 2},
  year      = {2026},
  pages        = {1268--1284},
  publisher    = {ACM},
  doi          = {10.1145/3779212.3790192},
  url          = {https://doi.org/10.1145/3779212.3790192}
}

@inproceedings{neupims,
  author    = {Guseul Heo and
               Sangyeop Lee and
               Jaehong Cho and
               Hyunmin Choi and
               Sanghyeon Lee and
               Hyungkyu Ham and
               Gwangsun Kim and
               Divya Mahajan and
               Jongse Park},
  editor    = {Rajiv Gupta and
               Nael B. Abu{-}Ghazaleh and
               Madan Musuvathi and
               Dan Tsafrir},
  title     = {{NeuPIMs}: {NPU-PIM} Heterogeneous Acceleration for Batched {LLM} Inferencing},
  booktitle = {Proceedings of the 29th {ACM} International Conference on Architectural
               Support for Programming Languages and Operating Systems, Volume 3,
               {ASPLOS} 2024, La Jolla, CA, USA, 27 April 2024- 1 May 2024},
  pages     = {722--737},
  publisher = {{ACM}},
  year      = {2024},
  url       = {https://doi.org/10.1145/3620666.3651380},
  doi       = {10.1145/3620666.3651380},
  bibsource = {dblp computer science bibliography, https://dblp.org}
}

@inproceedings{paise,
  author    = {Hyojung Lee and
               Daehyeon Baek and
               Jimyoung Son and
               Jieun Choi and
               Kihyo Moon and
               Minsung Jang},
  title     = {{PAISE:} {PIM}-Accelerated Inference Scheduling Engine for Transformer-based
               {LLM}},
  booktitle = {{IEEE} International Symposium on High Performance Computer Architecture,
               {HPCA} 2025, Las Vegas, NV, USA, March 1-5, 2025},
  pages     = {1707--1719},
  publisher = {{IEEE}},
  year      = {2025},
  url       = {https://doi.org/10.1109/HPCA61900.2025.00126},
  doi       = {10.1109/HPCA61900.2025.00126},
  bibsource = {dblp computer science bibliography, https://dblp.org}
}

@inproceedings{pimdl,
  author    = {Li, Cong and Zhou, Zhe and Wang, Yang and Yang, Fan and Cao, Ting and Yang, Mao and Liang, Yun and Sun, Guangyu},
  title     = {{PIM-DL}: Expanding the Applicability of Commodity {DRAM}-PIMs for Deep Learning via Algorithm-System Co-Optimization},
  year      = {2024},
  isbn      = {9798400703850},
  publisher = {Association for Computing Machinery},
  address   = {New York, NY, USA},
  url       = {https://doi.org/10.1145/3620665.3640376},
  doi       = {10.1145/3620665.3640376},
  booktitle = {Proceedings of the 29th ACM International Conference on Architectural Support for Programming Languages and Operating Systems, Volume 2},
  pages     = {879–896},
  numpages  = {18},
  location  = {La Jolla, CA, USA},
  series    = {ASPLOS '24}
}

@article{pimscalability,
  author    = {Gilbert Jonatan and
               Haeyoon Cho and
               Hyojun Son and
               Xiangyu Wu and
               Neal Livesay and
               Evelio Mora and
               Kaustubh Shivdikar and
               Jos{\'{e}} L. Abell{\'{a}}n and
               Ajay Joshi and
               David R. Kaeli and
               John Kim},
  title     = {Scalability Limitations of Processing-in-Memory using Real System
               Evaluations},
  journal   = {Proc. {ACM} Meas. Anal. Comput. Syst.},
  volume    = {8},
  number    = {1},
  pages     = {5:1--5:28},
  year      = {2024},
  url       = {https://doi.org/10.1145/3639046},
  doi       = {10.1145/3639046},
  bibsource = {dblp computer science bibliography, https://dblp.org}
}

@article{pimsurvey,
  title    = {A survey on processing-in-memory techniques: Advances and challenges},
  journal  = {Memories - Materials, Devices, Circuits and Systems},
  volume   = {4},
  pages    = {100022},
  year     = {2023},
  issn     = {2773-0646},
  doi      = {10.1016/j.memori.2022.100022},
  url      = {https://www.sciencedirect.com/science/article/pii/S2773064622000160},
  author   = {Kazi Asifuzzaman and Narasinga Rao Miniskar and Aaron R. Young and Frank Liu and Jeffrey S. Vetter}
}

@inproceedings{pimsys,
  author    = {Derek Christ and
               Lukas Steiner and
               Matthias Jung and
               Norbert Wehn},
  title     = {{PIMSys}: {A} Virtual Prototype for Processing in Memory},
  booktitle = {Proceedings of the International Symposium on Memory Systems, {MEMSYS}
               2024, Washington, DC, USA, 30 September 2024 - 3 October 2024},
  pages     = {26--33},
  publisher = {{ACM}},
  year      = {2024},
  url       = {https://doi.org/10.1145/3695794.3695797},
  doi       = {10.1145/3695794.3695797},
  bibsource = {dblp computer science bibliography, https://dblp.org}
}

@inproceedings{queen,
  author    = {Sharath Girish and
               Tianye Li and
               Amrita Mazumdar and
               Abhinav Shrivastava and
               David Luebke and
               Shalini De Mello},
  editor    = {Amir Globerson and Lester Mackey and Danielle Belgrave and Angela Fan and Ulrich Paquet and Jakub M. Tomczak and Cheng Zhang},
  title     = {{QUEEN:} QUantized Efficient ENcoding of Dynamic {Gaussians} for Streaming
               Free-viewpoint Videos},
  booktitle = {Advances in Neural Information Processing Systems},
  year      = {2024},
  url       = {https://proceedings.neurips.cc/paper_files/paper/2024/hash/4c9477b9e2c7ec0ad3f4f15077aaf85a-Abstract-Conference.html},
  bibsource = {dblp computer science bibliography, https://dblp.org},
  volume       = {37},
  pages        = {43435--43467},
  doi          = {10.52202/079017-1376}
}

@inproceedings{realtime3dhr,
  author    = {Yuanwang Yang and
               Qiao Feng and
               Yu{-}Kun Lai and
               Kun Li},
  editor    = {Takeo Igarashi and
               Ruizhen Hu},
  title     = {Real-time {3D} Human Reconstruction and Rendering System from a Single
{RGB} Camera},
  booktitle = {{SIGGRAPH} Asia 2024 Technical Communications, {SA} 2024, Tokyo, Japan,
               December 3-6, 2024},
  pages     = {6:1--6:4},
  publisher = {{ACM}},
  year      = {2024},
  url       = {https://doi.org/10.1145/3681758.3697993},
  doi       = {10.1145/3681758.3697993},
  bibsource = {dblp computer science bibliography, https://dblp.org}
}

@inproceedings{scaffoldgs,
  author    = {Tao Lu and
               Mulin Yu and
               Linning Xu and
               Yuanbo Xiangli and
               Limin Wang and
               Dahua Lin and
               Bo Dai},
  title     = {{Scaffold-GS}: Structured {3D} {Gaussians} for View-Adaptive Rendering},
  booktitle = {{IEEE/CVF} Conference on Computer Vision and Pattern Recognition,
               {CVPR} 2024, Seattle, WA, USA, June 16-22, 2024},
  pages     = {20654--20664},
  publisher = {{IEEE}},
  year      = {2024},
  url       = {https://doi.org/10.1109/CVPR52733.2024.01952},
  doi       = {10.1109/CVPR52733.2024.01952},
  bibsource = {dblp computer science bibliography, https://dblp.org}
}

@inproceedings{specpim,
  author    = {Cong Li and
               Zhe Zhou and
               Size Zheng and
               Jiaxi Zhang and
               Yun Liang and
               Guangyu Sun},
  editor    = {Rajiv Gupta and
               Nael B. Abu{-}Ghazaleh and
               Madan Musuvathi and
               Dan Tsafrir},
  title     = {{SpecPIM}: Accelerating Speculative Inference on {PIM}-Enabled System
via Architecture-Dataflow Co-Exploration},
  booktitle = {Proceedings of the 29th {ACM} International Conference on Architectural
               Support for Programming Languages and Operating Systems, Volume 3,
               {ASPLOS} 2024, La Jolla, CA, USA, 27 April 2024- 1 May 2024},
  pages     = {950--965},
  publisher = {{ACM}},
  year      = {2024},
  url       = {https://doi.org/10.1145/3620666.3651352},
  doi       = {10.1145/3620666.3651352},
  bibsource = {dblp computer science bibliography, https://dblp.org}
}

@inproceedings{stgfs,
  author    = {Zhan Li and
               Zhang Chen and
               Zhong Li and
               Yi Xu},
  title     = {Spacetime {Gaussian} Feature Splatting for Real-Time Dynamic View Synthesis},
  booktitle = {{IEEE/CVF} Conference on Computer Vision and Pattern Recognition,
               {CVPR} 2024, Seattle, WA, USA, June 16-22, 2024},
  pages     = {8508--8520},
  publisher = {{IEEE}},
  year      = {2024},
  url       = {https://doi.org/10.1109/CVPR52733.2024.00813},
  doi       = {10.1109/CVPR52733.2024.00813},
  bibsource = {dblp computer science bibliography, https://dblp.org}
}

@article{temporalcoherence,
  author    = {Daniel Scherzer and
               Lei Yang and
               Oliver Mattausch and
               Diego Nehab and
               Pedro V. Sander and
               Michael Wimmer and
               Elmar Eisemann},
  title     = {Temporal Coherence Methods in Real-Time Rendering},
  journal   = {Comput. Graph. Forum},
  volume    = {31},
  number    = {8},
  pages     = {2378--2408},
  year      = {2012},
  url       = {https://doi.org/10.1111/j.1467-8659.2012.03075.x},
  doi       = {10.1111/J.1467-8659.2012.03075.X},
  bibsource = {dblp computer science bibliography, https://dblp.org}
}

@inproceedings{thudgs,
  author    = {Xiaolong Yang and
               Yang Wang and
               Wende Xu and
               Yubin Qin and
               Huanyu Wang and
               Ruiqi Guo and
               Zhiheng Yue and
               Jiangyuan Gu and
               Shaojun Wei and
               Yang Hu and
               Shouyi Yin},
  title     = {2.9 {A} 0.24{mJ/Frame} Quadratic Interpolation {4DGS} Processor with Recursive
Computation Reuse and Tree-Based Parallel-Rendering},
  booktitle = {{IEEE} International Solid-State Circuits Conference, {ISSCC} 2026,
               San Francisco, CA, USA, February 15-19, 2026},
  pages     = {58--60},
  publisher = {{IEEE}},
  year      = {2026},
  url       = {https://doi.org/10.1109/ISSCC49663.2026.11409230},
  doi       = {10.1109/ISSCC49663.2026.11409230},
  bibsource = {dblp computer science bibliography, https://dblp.org}
}

@inproceedings{wu4dgs,
  author    = {Guanjun Wu and
               Taoran Yi and
               Jiemin Fang and
               Lingxi Xie and
               Xiaopeng Zhang and
               Wei Wei and
               Wenyu Liu and
               Qi Tian and
               Xinggang Wang},
  title     = {{4D} {Gaussian} Splatting for Real-Time Dynamic Scene Rendering},
  booktitle = {{IEEE/CVF} Conference on Computer Vision and Pattern Recognition,
               {CVPR} 2024, Seattle, WA, USA, June 16-22, 2024},
  pages     = {20310--20320},
  publisher = {{IEEE}},
  year      = {2024},
  url       = {https://doi.org/10.1109/CVPR52733.2024.01920},
  doi       = {10.1109/CVPR52733.2024.01920},
  bibsource = {dblp computer science bibliography, https://dblp.org}
}

@inproceedings{yang4dgs,
  author    = {Zeyu Yang and
               Hongye Yang and
               Zijie Pan and
               Li Zhang},
  title     = {Real-time Photorealistic Dynamic Scene Representation and Rendering
with {4D} {Gaussian} Splatting},
  booktitle = {The Twelfth International Conference on Learning Representations,
               {ICLR} 2024, Vienna, Austria, May 7-11, 2024},
  publisher = {OpenReview.net},
  year      = {2024},
  url       = {https://openreview.net/forum?id=WhgB5sispV},
  bibsource = {dblp computer science bibliography, https://dblp.org}
}

@article{tanksandtemples,
  author  = {Arno Knapitsch and Jaesik Park and Qian{-}Yi Zhou and Vladlen Koltun},
  title   = {{Tanks and Temples}: Benchmarking Large-Scale Scene Reconstruction},
  journal = {ACM Transactions on Graphics},
  volume  = {36},
  number  = {4},
  pages   = {78:1--78:13},
  year    = {2017},
  doi     = {10.1145/3072959.3073599}
}

@inproceedings{sarangi2021deepscaletool,
  author    = {Sarangi, Satyabrata and Baas, Bevan M.},
  title     = {{DeepScaleTool}: A Tool for the Accurate Estimation of Technology Scaling in the Deep-Submicron Era},
  booktitle = {2021 IEEE International Symposium on Circuits and Systems (ISCAS)},
  year      = {2021},
  pages     = {1--5},
  doi       = {10.1109/ISCAS51556.2021.9401196}
}

@inproceedings{gscore,
  author    = {Junseo Lee and Seokwon Lee and Jungi Lee and Junyong Park and Jaewoong Sim},
  title     = {{GSCore}: Efficient Radiance Field Rendering via Architectural Support for {3D} {Gaussian} Splatting},
  booktitle = {Proceedings of the 29th ACM International Conference on Architectural Support for Programming Languages and Operating Systems},
  pages     = {497--511},
  publisher = {ACM},
  year      = {2024},
  doi       = {10.1145/3620666.3651385}
}

@inproceedings{gstg,
  author    = {Joongho Jo and Jongsun Park},
  title     = {{GS-TG}: {3D} {Gaussian} Splatting Accelerator with Tile Grouping for Reducing Redundant Sorting while Preserving Rasterization Efficiency},
  booktitle = {Proceedings of the 62nd ACM/IEEE Design Automation Conference},
  pages     = {1--7},
  publisher = {IEEE},
  year      = {2025},
  doi       = {10.1109/DAC63849.2025.11133283}
}

@inproceedings{gaupre,
  author    = {Yuzheng Lin and Lizhou Wu and Chixiao Chen and Xiaoyang Zeng and Haozhe Zhu},
  title     = {{GauPRE}: A Pattern-based Rendering Engine for {Gaussian} Splatting on Edge Device},
  booktitle = {Proceedings of the IEEE/ACM International Conference on Computer-Aided Design},
  pages     = {1--9},
  publisher = {IEEE},
  year      = {2025},
  doi       = {10.1109/ICCAD66269.2025.11240948}
}

@inproceedings{lumina,
  author    = {Yu Feng and Weikai Lin and Yuge Cheng and Zihan Liu and Jingwen Leng and Minyi Guo and Chen Chen and Shixuan Sun and Yuhao Zhu},
  title     = {{Lumina}: Real-Time Neural Rendering by Exploiting Computational Redundancy},
  booktitle = {Proceedings of the 52nd Annual International Symposium on Computer Architecture},
  pages     = {1925--1939},
  publisher = {ACM},
  year      = {2025},
  doi       = {10.1145/3695053.3731003}
}

@inproceedings{orange,
  author    = {Haomin Li and Yue Liang and Fangxin Liu and Bowen Zhu and Zongwu Wang and Yu Feng and Liqiang Lu and Li Jiang and Haibing Guan},
  title     = {{ORANGE}: Exploring Ockham's Razor for Neural Rendering by Accelerating {3DGS} on {NPUs} with {GEMM}-Friendly Blending and Balanced Workloads},
  booktitle = {Proceedings of the IEEE International Symposium on High Performance Computer Architecture},
  pages     = {1--15},
  publisher = {IEEE},
  year      = {2026},
  doi       = {10.1109/HPCA68181.2026.11408575}
}

@inproceedings{grtx,
  author    = {Junseo Lee and Sangyun Jeon and Jungi Lee and Junyong Park and Jaewoong Sim},
  title     = {{GRTX}: Efficient Ray Tracing for {3D} {Gaussian}-Based Rendering},
  booktitle = {Proceedings of the IEEE International Symposium on High Performance Computer Architecture},
  pages     = {1--14},
  publisher = {IEEE},
  year      = {2026},
  doi       = {10.1109/HPCA68181.2026.11408531}
}

@inproceedings{nebula,
  author    = {He Zhu and Zheng Liu and Xingyang Li and Anbang Wu and Jieru Zhao and Fangxin Liu and Yiming Gan and Jingwen Leng and Yu Feng},
  title     = {Nebula: Infinite-Scale {3D} {Gaussian} Splatting in {VR} via Collaborative Rendering and Accelerated Stereo Rasterization},
  booktitle = {Proc. 31st {ACM ASPLOS}},
  pages     = {1235--1250},
  publisher = {ACM},
  year      = {2026},
  doi       = {10.1145/3779212.3790190}
}

@inproceedings{gsarch,
  author    = {Houshu He and Gang Li and Fangxin Liu and Li Jiang and Xiaoyao Liang and Zhuoran Song},
  title     = {{GSArch}: Breaking Memory Barriers in {3D} {Gaussian} Splatting Training via Architectural Support},
  booktitle = {Proceedings of the IEEE International Symposium on High Performance Computer Architecture},
  pages     = {366--379},
  publisher = {IEEE},
  year      = {2025},
  doi       = {10.1109/HPCA61900.2025.00037}
}

@inproceedings{gauspu,
  author    = {Lizhou Wu and Haozhe Zhu and Siqi He and Jiapei Zheng and Chixiao Chen and Xiaoyang Zeng},
  title     = {{GauSPU}: {3D} {Gaussian} Splatting Processor for Real-Time {SLAM} Systems},
  booktitle = {Proceedings of the 57th IEEE/ACM International Symposium on Microarchitecture},
  pages     = {1562--1573},
  publisher = {IEEE},
  year      = {2024},
  doi       = {10.1109/MICRO61859.2024.00114}
}

@inproceedings{react3d,
  author    = {Hongyi Wang and Zhenhua Zhu and Tianchen Zhao and Yunfei Xiang and Zehao Wang and Jincheng Yu and Huazhong Yang and Yuan Xie and Yu Wang},
  title     = {{REACT3D}: Real-Time Edge Accelerator for Incremental Training in {3D} {Gaussian} Splatting Based {SLAM} Systems},
  booktitle = {Proceedings of the 58th IEEE/ACM International Symposium on Microarchitecture},
  pages     = {1852--1866},
  publisher = {ACM},
  year      = {2025},
  doi       = {10.1145/3725843.3756109}
}

@inproceedings{cambricongs,
  author    = {Rui Wen and Zhifei Yue and Tianbo Liu and Xinkai Song and Jin Li and Di Huang and Jiaming Guo and Xing Hu and Zidong Du and Qi Guo and Tianshi Chen},
  title     = {{Cambricon-GS}: An Accelerator for {3D} {Gaussian} Splatting Training with {Gaussian}-Pixel Hybrid Parallelism},
  booktitle = {Proceedings of the IEEE International Symposium on High Performance Computer Architecture},
  pages     = {1--14},
  publisher = {IEEE},
  year      = {2026},
  doi       = {10.1109/HPCA68181.2026.11408562}
}

@inproceedings{splatonic,
  author    = {Xiaotong Huang and He Zhu and Tianrui Ma and Yuxiang Xiong and Fangxin Liu and Zhezhi He and Yiming Gan and Zihan Liu and Jingwen Leng and Yu Feng and Minyi Guo},
  title     = {{SPLATONIC}: Architectural Support for {3D} {Gaussian} Splatting {SLAM} via Sparse Processing},
  booktitle = {Proceedings of the IEEE International Symposium on High Performance Computer Architecture},
  pages     = {1--14},
  publisher = {IEEE},
  year      = {2026},
  doi       = {10.1109/HPCA68181.2026.11408505}
}

@inproceedings{ags,
  author    = {Houshu He and Naifeng Jing and Li Jiang and Xiaoyao Liang and Zhuoran Song},
  title     = {{AGS}: Accelerating {3D} {Gaussian} Splatting {SLAM} via {CODEC}-Assisted Frame Covisibility Detection},
  booktitle = {Proceedings of the 31st ACM International Conference on Architectural Support for Programming Languages and Operating Systems},
  pages     = {20--34},
  publisher = {ACM},
  year      = {2026},
  doi       = {10.1145/3760250.3762229}
}

@inproceedings{relaygs,
  author    = {Yeonggeon Kim and Seongmin Ki and Sangkyu Jeon and Sungju Ryu},
  title     = {{Relay-GS}: Reusing Temporal Sort Information for {4D} {Gaussian} Splatting Acceleration},
  booktitle = {Proceedings of the 63rd ACM/IEEE Design Automation Conference},
  publisher = {ACM},
  year      = {2026},
  pages        = {1--7},
  doi          = {10.1145/3770743.3804261},
  url          = {https://doi.org/10.1145/3770743.3804261}
}

@inproceedings{specter,
  author    = {Jae-Young Kim and Teokkyu Suh and Junsoo Lim and Minjun Park and Hyunmin Choi and Jaeha Min and Donghyuk Kim and Joo-Young Kim},
  title     = {{Specter}: 376 fps 2.67 {nJ/Pixel} {4DGS} Processor for Spatiotemporal Reconstruction with Contribution-Ordered Rendering and Deformation-Aware Modeling},
  booktitle = {Proceedings of the IEEE/JSAP Symposium on VLSI Technology and Circuits},
  pages     = {1--3},
  publisher = {IEEE},
  year      = {2026},
  doi       = {10.1109/VLSITechnologyandCir65830.2026.11577601}
}

@inproceedings{flicker,
  author    = {Wenhui Ou and Zhuoyu Wu and Yipu Zhang and Dongjun Wu and Freddy Ziyang Hong and C. Patrick Yue},
  title     = {{FLICKER}: A Fine-Grained Contribution-Aware Accelerator for Real-Time {3D} {Gaussian} Splatting},
  booktitle = {Proceedings of the Design, Automation and Test in Europe Conference},
  pages     = {1--7},
  publisher = {IEEE},
  year      = {2026},
  doi       = {10.23919/DATE69613.2026.11539442}
}

@article{chang129fps,
  author    = {Fang-Chi Chang and Tian-Sheuan Chang},
  title     = {A 129 {FPS} {Full HD} Real-Time Accelerator for {3D} {Gaussian} Splatting},
  journal   = {IEEE Transactions on Visualization and Computer Graphics},
  volume    = {32},
  number    = {7},
  pages     = {6537--6547},
  year      = {2026},
  doi       = {10.1109/TVCG.2026.3683714}
}

@inproceedings{localgs,
  author    = {Yiyang Sun and Qinzhe Zhi and Yiqi Jing and Le Ye and Ru Huang and Tianyu Jia},
  title     = {{Local-GS}: An Order-Independent {Gaussian} Splatting Training Accelerator Exploiting Splat Locality},
  booktitle = {Proceedings of the 62nd ACM/IEEE Design Automation Conference},
  pages     = {1--7},
  publisher = {IEEE},
  year      = {2025},
  doi       = {10.1109/DAC63849.2025.11132792}
}

@inproceedings{boa3dgs,
  author    = {Hyukjun Kweon and Jongyeop Kim and Jeongwoo Park},
  title     = {{BOA-3DGS}: Backward-Striding Optimized Accelerator for Reduced Memory Contention in {3D} {Gaussian} Splatting Training},
  booktitle = {Proceedings of the 31st Asia and South Pacific Design Automation Conference},
  pages     = {341--347},
  publisher = {IEEE},
  year      = {2026},
  doi       = {10.1109/ASP-DAC66049.2026.11420496}
}

@inproceedings{he2025lpspec,
  author       = {Siyuan He and Zhantong Zhu and Yandong He and Tianyu Jia},
  title        = {{LP-Spec}: Leveraging {LPDDR PIM} for Efficient {LLM} Mobile Speculative Inference with Architecture-Dataflow Co-Optimization},
  year         = {2025},
  url          = {https://doi.org/10.1109/ICCAD66269.2025.11240889},
  booktitle    = {2025 IEEE/ACM International Conference on Computer Aided Design (ICCAD)},
  pages        = {1--9},
  publisher    = {IEEE},
  doi          = {10.1109/ICCAD66269.2025.11240889}
}

@misc{samsung2017eightlpp,
  author       = {{Samsung Electronics}},
  title        = {Samsung Completes Qualification of {8nm LPP} Process},
  year         = {2017},
  howpublished = {Samsung Global Newsroom},
  url          = {https://news.samsung.com/global/samsung-completes-qualification-of-8nm-lpp-process},
  month        = {October}
}

@manual{micron2021lpddr5,
  author       = {{Micron Technology, Inc.}},
  title        = {{LPDDR5 SDRAM}},
  year         = {2021},
  month        = may,
  note         = {y31m\_lpddr5\_multi\_pkg.pdf, Rev. A},
  url          = {https://mm.digikey.com/Volume0/opasdata/d220001/medias/docus/8885/441b-315b-y31m-lpddr5-multi-pkg.pdf}
}

@inproceedings{waymo,
  author    = {Pei Sun and Henrik Kretzschmar and Xerxes Dotiwalla and Aurelien Chouard and Vijaysai Patnaik and Paul Tsui and James Guo and Yin Zhou and Yuning Chai and Benjamin Caine and Vijay Vasudevan and Wei Han and Jiquan Ngiam and Hang Zhao and Aleksei Timofeev and Scott Ettinger and Maxim Krivokon and Amy Gao and Aditya Joshi and Yu Zhang and Jonathon Shlens and Zhifeng Chen and Dragomir Anguelov},
  title     = {Scalability in Perception for Autonomous Driving: {Waymo} Open Dataset},
  booktitle = {Proceedings of the IEEE/CVF Conference on Computer Vision and Pattern Recognition (CVPR)},
  year      = {2020},
  pages        = {2446--2454},
  doi          = {10.1109/CVPR42600.2020.00252},
  url          = {https://openaccess.thecvf.com/content_CVPR_2020/html/Sun_Scalability_in_Perception_for_Autonomous_Driving_Waymo_Open_Dataset_CVPR_2020_paper.html}
}

@inproceedings{matrixcity,
  author    = {Yixuan Li and Lihan Jiang and Linning Xu and Yuanbo Xiangli and Zhenzhi Wang and Dahua Lin and Bo Dai},
  title     = {{MatrixCity}: A Large-scale City Dataset for City-scale Neural Rendering and Beyond},
  booktitle = {Proceedings of the IEEE/CVF International Conference on Computer Vision (ICCV)},
  year      = {2023},
  pages     = {3205--3215},
  doi       = {10.1109/ICCV51070.2023.00297},
  url       = {https://openaccess.thecvf.com/content/ICCV2023/html/Li_MatrixCity_A_Large-scale_City_Dataset_for_City-scale_Neural_Rendering_and_ICCV_2023_paper.html}
}

@article{ramulator2,
  author    = {Haocong Luo and Yahya Can Tu{\u{g}}rul and F. Nisa Bostanc{\i} and Ataberk Olgun and A. Giray Ya{\u{g}}l{\i}k{\c{c}}{\i} and Onur Mutlu},
  title     = {{Ramulator 2.0}: A Modern, Modular, and Extensible {DRAM} Simulator},
  journal   = {IEEE Computer Architecture Letters},
  volume    = {23},
  number    = {1},
  pages     = {112--116},
  year      = {2024},
  doi       = {10.1109/LCA.2023.3333759},
  url       = {https://doi.org/10.1109/LCA.2023.3333759}
}

@inproceedings{pei2026degs,
  author    = {Minnan Pei and Gang Li and Zeyu Zhu and
               Siting Wang and Junwen Si and Zhuoran Song and
               Yu Feng and Fangxin Liu and Xiaoyao Liang and
               Jian Cheng},
  title     = {{DeGS}: A Scalable {3DGS} Architecture via
               Decoupled Workload Parsing and Reorganization},
  booktitle = {2026 59th IEEE/ACM International Symposium
               on Microarchitecture (MICRO)},
  year      = {2026},
  pages     = {848--862},
  publisher = {IEEE},
  doi       = {10.1109/MICRO67795.2026.00070},
  url       = {https://doi.org/10.1109/MICRO67795.2026.00070}
}
%%%%%%%%%%%%%%%%%%%%%%%%%%%%%%%%%%%%

% Artifact documentation begins on a separate page after the references.
\clearpage
\begingroup
\renewcommand{\encodingdefault}{T1}
\normalfont
\lstset{basicstyle=\ttfamily\footnotesize,columns=fullflexible,
  keepspaces=true,showstringspaces=false,breaklines=true,upquote=true,
  aboveskip=4pt,belowskip=4pt}
\clubpenalty=10000
\widowpenalty=10000
\widowpenalties 3 10000 10000 0
\displaywidowpenalty=10000
% Adapted from the cTuning Artifact Appendix template V20240722.
% Template copyright 2014--2024 cTuning.org, CC BY 4.0.
% https://github.com/ctuning/artifact-evaluation/blob/master/docs/template/ae.tex
% Use with the paper's IEEEtran class, hyperref, booktabs and listings.
\appendix[Artifact Appendix]

\subsection{Abstract}
This artifact reproduces the speedup comparison in Figure~\ref{fig:overall-performance}, the ablation study in Figure~\ref{fig:ablation}, and the rendering quality in Table~\ref{tab:quality-gpu} of GSPIM. It includes CUDA rendering, GPU execution capture, trace construction, a Ramulator2-based GPU/PIM scheduling model, released checkpoints, and report generation. Evaluation covers 4DGS-1K, Ex4DGS, and Anchored 4DGS on six scenes. GPU execution is measured on Jetson AGX Orin. PIM and memory execution are simulated using captured GPU events and physical memory requests.

\subsection{Artifact check-list (meta-information)}
{\small
\begin{itemize}
\item \textbf{Algorithm:} Temporal selection, multi-frame rendering, data reorganization, and overlapping GPU/PIM scheduling.
\item \textbf{Program:} Python, C++, and CUDA. Chisel hardware and model-training sources are also included.
\item \textbf{Compilation:} GCC/G++ 13.3, CMake, Ninja, and CUDA 13.2. Native components build automatically.
\item \textbf{Model:} 18 released checkpoints for three models.
\item \textbf{Data set:} Coffee Martini, Cook Spinach, Sear Steak, Flame Steak, Dance3, and Actor1\_4.
\item \textbf{Run-time environment:} Ubuntu 24.04 / L4T R39.2, Python 3.12, and PyTorch 2.14.0+cu132.
\item \textbf{Hardware:} Jetson AGX Orin 64 GB, aarch64, compute capability 8.7, and an external SSD.
\item \textbf{Run-time state:} MAXN, platform frequency governor without clock locking, and no concurrent GPU workloads.
\item \textbf{Metrics and output:} Speedups, PSNR, SSIM, Markdown tables, and PNG/PDF figures.
\item \textbf{Disk space:} Approximately 100 GB, including about 48 GB of prepared data and 4 GB of extracted checkpoints.
\item \textbf{Preparation time:} About 44 min with local raw inputs. Downloads, installation, and initial compilation add time.
\item \textbf{Experiment time:} About 3 h 33 min for performance with fresh traces/schedules and compiled extensions, then 24 min for full-sequence quality using those caches.
\item \textbf{Publicly available:} Source and checkpoints.
\item \textbf{Licenses:} GSPIM source is MIT. Third-party code, models, and data retain their terms in \path{LICENSES/THIRD_PARTY.md}.
\item \textbf{Workflow:} \texttt{python -m gspim.demo}.
\item \textbf{Archived:} \url{https://doi.org/10.5281/zenodo.21447096}.
\end{itemize}}

\subsection{Description}
\subsubsection{How to access}
The Zenodo record contains source v1.0, four checkpoint archives v1.1, and \texttt{SHA256SUMS} (3.74 GB total). Source v1.0 corresponds to \href{https://github.com/MAdrid1011/GSPIM/tree/ef5191b86f325c943af220b74e8a3f4fcb33e232}{commit \texttt{ef5191b86f32}}, with the full revision in \texttt{ARCHIVE.md}. The \href{https://github.com/MAdrid1011/GSPIM}{project repository} also provides development source and the \href{https://github.com/MAdrid1011/GSPIM/releases/tag/checkpoints-v1.1}{checkpoint release}. Preserve the bundled Git metadata used to verify pinned dependencies. Input hashes and upstream revisions are recorded in \path{integrations/manifest.py}.

\subsubsection{Hardware and software dependencies}
Use the Orin platform and software stack above with CUDA and Conda installed. Python dependencies include gsplat 1.5.3, NumPy 2.0.1, Pillow 12.1.1, OpenCV 5.0.0.93, and Matplotlib. Conda-forge FFmpeg 5.1.0 provides H.264/HEVC software decoding. Package installation needs network access and system packages need administrator privileges. Preparation initializes pinned Ramulator2 and Golden-Ratio sources.

\subsubsection{Data sets and models}
Preparation downloads the four Neural 3D Video scenes. Request Dance3 and Actor1\_4 through the Mobile-Stage and ENeRF-Outdoor forms in \path{docs/data.md}. Released checkpoints cover all 18 model/scene combinations. Held-out cameras are 00 for Neural 3D Video and Actor1\_4, and 04 for Dance3. Prepared images are $1352\times1014$ and $960\times540$, respectively. Representation-specific resizing and camera conventions follow \path{docs/data.md} in the archived source.

\subsection{Installation}
Download all five archives and \texttt{SHA256SUMS} from Zenodo into one directory. In Bash, install the build tools, verify every archive, and extract the source:
\begin{lstlisting}
sudo apt-get install build-essential cmake \
  ninja-build git zstd
sha256sum -c SHA256SUMS
tar --zstd -xf gspim-ae-v1.0-source.tar.zst
export GSPIM_STORAGE=/mnt/ssd/gspim-ae
mkdir -p "$GSPIM_STORAGE"/{datasets,checkpoints,\
cache/downloads,cache/tmp,results}
for a in gspim-checkpoints-v1.1-*.tar.zst; do
  tar --zstd -xf "$a" -C "$GSPIM_STORAGE"
done
cd GSPIM
conda create --prefix "$GSPIM_STORAGE/env" \
  --override-channels -c conda-forge \
  python=3.12 pip ffmpeg=5.1.0 -y
conda activate "$GSPIM_STORAGE/env"
export CUDA_HOME=/usr/local/cuda
export PATH="$CUDA_HOME/bin:$PATH"
export TMPDIR="$GSPIM_STORAGE/cache/tmp"
python -m pip install torch==2.14.0 \
  --index-url https://download.pytorch.org/whl/cu132
python -m pip install -e '.[cuda]'
python -m pip install \
  opencv-python-headless==5.0.0.93
ln -s "$GSPIM_STORAGE/datasets" datasets
ln -s "$GSPIM_STORAGE/checkpoints" checkpoints
\end{lstlisting}
Place the two requested scene archives in this cache directory:
\begin{lstlisting}
$GSPIM_STORAGE/cache/downloads/
  dance3_libx265.tar.gz
  actor1_4_libx265.tar.gz
\end{lstlisting}
Prepare inputs and check checkpoints from the repository root:
\begin{lstlisting}
python -m gspim.demo --prepare \
  --cache "$GSPIM_STORAGE/cache"
\end{lstlisting}

\subsection{Experiment workflow}
\subsubsection{Figure~\ref{fig:overall-performance} speedups and Figure~\ref{fig:ablation} ablation}
The run uses the first 10 consecutive held-out frames, 100 warmup windows, and four measurement repetitions per combination. CUDA events measure Baseline and Dataflow, alternating their measurement order. Timing excludes loading, compilation, preprocessing, and quality evaluation.
\begin{lstlisting}
python -m gspim.demo \
  --cache "$GSPIM_STORAGE/cache" \
  --out "$GSPIM_STORAGE/results/performance"
\end{lstlisting}

PIM-only uses independent-frame traces. Serial combines Dataflow and PIM with batch/window completion barriers. Full overlaps preparation and GPU work subject to producer dependencies and buffer ownership. Speedup is the ratio of Baseline mean per-frame time to the measured or simulated mean per-frame time of each configuration.

\subsubsection{Table~\ref{tab:quality-gpu} quality}
The run evaluates 300 frames per scene and 200 for Actor1\_4, totaling 5,100 frames across all models. PSNR and RGB SSIM are computed per frame against held-out images and averaged per model/scene. SSIM uses an $11\times11$ Gaussian window with standard deviation 1.5.
\begin{lstlisting}
python -m gspim.demo --frames 300 \
  --cache "$GSPIM_STORAGE/cache" \
  --out "$GSPIM_STORAGE/results/quality"
\end{lstlisting}

This pass reuses the 10-frame PIM traces/schedules and collects fresh full-sequence GPU measurements and quality metrics. Full quality uses Dataflow-rendered images because modeled PIM operations select and reorganize records. Captured-frame RGB hashes and window decisions are checked against current rendering before evaluation.

\subsection{Evaluation and expected results}
Both runs produce \texttt{verification.md}, per-combination \texttt{summary.md} files, and \texttt{performance.\{png,pdf\}} and \texttt{ablation.\{png,pdf\}}. Use the performance directory for Figures~\ref{fig:overall-performance} and~\ref{fig:ablation}, and the quality section of the quality report for Table~\ref{tab:quality-gpu}. Progress appears in the terminal. Traces, schedules, and raw measurements remain in the external cache.

Figure~\ref{fig:overall-performance} compares Baseline, Dataflow, and Full. Figure~\ref{fig:ablation} also includes PIM-only and Serial. Reference geometric-mean speedups over all 18 combinations are:
\begin{center}
\begin{tabular}{rrrrr}
\toprule
Baseline & Dataflow & PIM-only & Serial & Full \\
\midrule
$1.00\times$ & $2.40\times$ & $3.19\times$ & $4.10\times$ & $5.36\times$ \\
\bottomrule
\end{tabular}
\end{center}
Compare the 72 nonbaseline model/scene speedups with the paper. The archived README gives a $\pm5\%$ relative speedup tolerance for this shorter run against the paper results. Table~\ref{tab:quality-gpu} has 108 quality values. Maximum scene-mean Dataflow/Full decreases in the reference run are 0.0059251 dB PSNR and 0.0000194526 SSIM. Reports show three PSNR decimals and four SSIM decimals. Compare with Table~\ref{tab:quality-gpu} at this precision, where differences peak at 0.006 dB and 0.0001.

\subsection{Experiment customization}
Use \texttt{-{}-workload 4dgs1k}, \texttt{ex4dgs}, or \texttt{anchored}, and a \texttt{-{}-scene} selector. A single-combination check adds \texttt{-{}-workload anchored -{}-scene dance3} with a separate output directory. Compatible caches are reused. GPU measurements and quality are collected on each invocation. \texttt{-{}-repeats} and \texttt{-{}-warmup} control sampling.

\endgroup

\end{document}